\documentclass[sigconf]{acmart}

\copyrightyear{2024}
\acmYear{2024}
\setcopyright{rightsretained}
\acmBooktitle{Proceedings of the 2024 Workshop on Attacks and Solutions in Hardware Security (ASHES '24), October 14--18, 2024, Salt Lake City, UT, USA}
\acmDOI{10.1145/3689939.3695779}
\acmISBN{979-8-4007-1235-7/24/10}
\acmConference[ASHES '24]{the 2024 Workshop on Attacks and Solutions in Hardware Security}{October 14, 2024}{Salt Lake City, USA}

\usepackage[nolist]{acronym}
\usepackage[linesnumbered,ruled,vlined]{algorithm2e}
\usepackage{algpseudocode}
\usepackage{booktabs}
\usepackage{pgf}
\usepackage{tikz}
\usetikzlibrary{arrows,automata}
\usepackage{enumitem}
\setlist[description]{leftmargin=0cm,labelindent=0cm}
\usepackage[font=footnotesize]{subcaption}
\usepackage{xcolor}

\newcommand{\circled}[2][]{
  \tikz[baseline=(char.base)]{
    \node[shape=circle,draw,inner sep=1pt,fill=white]
    (char) {\phantom{\ifblank{#1}{#2}{#1}}};
    \node[text=black] at (char.center) {\makebox[0pt][c]{\bfseries#2}};}\xspace}
\robustify{\circled}

\newcommand{\circledgreen}[2][]{
  \tikz[baseline=(char.base)]{
    \node[shape=circle,draw,inner sep=1pt,fill=green]
    (char) {\phantom{\ifblank{#1}{#2}{#1}}};
    \node[text=black] at (char.center) {\makebox[0pt][c]{\bfseries#2}};}\xspace}
\robustify{\circledgreen}

\newcommand{\circledred}[2][]{
	\tikz[baseline=(char.base)]{
		\node[shape=circle,inner sep=1pt,fill=lightcoral]
		(char) {\phantom{\ifblank{#1}{#2}{#1}}};
		\node[text=black] at (char.center) {\makebox[0pt][c]{\textbf#2}};}\xspace}
\robustify{\circledred}

\newcommand{\circledblue}[2][]{
	\tikz[baseline=(char.base)]{
		\node[shape=circle,inner sep=1pt,fill=lightskyblue]
		(char) {\phantom{\ifblank{#1}{#2}{#1}}};
		\node[text=black] at (char.center) {\makebox[0pt][c]{\textbf#2}};}\xspace}
\robustify{\circledblue}

\newcommand{\circledblack}[2][]{
	\tikz[baseline=(char.base)]{
		\node[shape=circle,inner sep=1pt,fill=black]
		(char) {\phantom{\ifblank{#1}{#2}{#1}}};
		\node[text=white] at (char.center) {\makebox[0pt][c]{\textbf#2}};}\xspace}
\robustify{\circledblack}

\definecolor{golden}{HTML}{E7A000}

\newcommand{\circledgold}[2][]{
  \tikz[baseline=(char.base)]{
    \node[shape=circle,draw,inner sep=1pt,color=golden]
    (char) {\phantom{\ifblank{#1}{#2}{#1}}};
    \node[text=golden] at (char.center) {\makebox[0pt][c]{\bfseries#2}};}\xspace}
\robustify{\circledgold}

\newcommand{\ie}{i.\,e.}
\newcommand{\eg}{e.\,g.}
\newcommand{\cf}{cf.\@\,}

\newcommand{\etal}{et~al.\@\,}

\makeatletter
\gdef\@copyrightpermission{
  \begin{minipage}{0.3\columnwidth}
   \href{https://creativecommons.org/licenses/by/4.0/}{\includegraphics[width=0.90\textwidth]{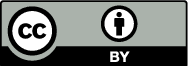}}
  \end{minipage}\hfill
  \begin{minipage}{0.7\columnwidth}
   \href{https://creativecommons.org/licenses/by/4.0/}{This work is licensed under a Creative Commons Attribution International 4.0 License.}
  \end{minipage}
  \vspace{5pt}
}
\makeatother

\begin{document}

\title[The Security Implications of Bitstream Modifications]{Patching FPGAs:\\The Security Implications of Bitstream Modifications}

\author{Endres Puschner}
\orcid{0000-0002-1661-6024}
\affiliation{%
  \institution{Max Planck Institute for Security and Privacy}
  \city{Bochum}
  \country{Germany}
}
\email{endres.puschner@mpi-sp.org}

\author{Maik Ender}
\orcid{0000-0002-0685-2541}
\affiliation{%
  \institution{Max Planck Institute for Security and Privacy}
  \city{Bochum}
  \country{Germany}
}
\email{maik.ender@mpi-sp.org}

\author{Steffen Becker}
\orcid{0000-0001-7526-5597}
\affiliation{%
  \institution{Ruhr University Bochum}
  \city{Bochum}
  \country{Germany}
}
\affiliation{%
  \institution{Max Planck Institute for Security and Privacy}
  \city{Bochum}
  \country{Germany}
}
\email{steffen.becker@rub.de}

\author{Christof Paar}
\orcid{0000-0001-8681-2277}
\affiliation{%
  \institution{Max Planck Institute for Security and Privacy}
  \city{Bochum}
  \country{Germany}
}
\email{christof.paar@mpi-sp.org}

\begin{acronym}
    \acro{AES}{Advanced Encryption Standard}
    \acro{AI}{Artificial Intelligence}
    \acro{API}{Application Programming Interface}
    \acro{ASIC}{Application-Specific Integrated Circuit}
    \acro{ATPG}{Automatic Test Pattern Generation}

    \acro{BEL}{Basic Element}
    \acro{BRAM}{Block Random-Access Memory}

    \acro{CLB}{Configurable Logic Block}
    \acro{CMOS}{Complementary Metal–Oxide–Semiconductor}
    \acro{CPU}{Central Processing Unit}
    \acro{CRC}{Cyclic Redundancy Checks}
    \acro{CRT}{Chinese Remainder Theorem}
    
    \acro{DL}{Deep Learning}
    \acro{DLL}{Dynamic Link Library}
    \acroplural{DLL}[DLLs]{Dynamic Link Libraries}
    \acro{DRC}{Design Rule Check}
    \acro{DRM}{Digital Rights Management}
    \acro{DSP}{Digital Signal Processor}
    
    \acro{ECC}{Elliptic-Curve Cryptography}
    \acro{ECO}{Engineering Change Order}
    \acro{EDA}{Electronic Design Automation}

    \acro{FASM}{FPGA Assembly}
    \acro{FF}{Flip Flop}
    \acro{FFT}{Fast Fourier Transformation}
    \acro{FIB}{Focused Ion Beam}
    \acro{FPGA}{Field Programmable Gate Array}
    
    \acro{GCD}{Greatest Common Divisor}
    \acro{GDSII}{Graphic Design System II}
    
    \acro{HDL}{Hardware Description Language}
    \acro{HSM}{Hardware Security Module}
    
    \acro{IC}{Integrated Circuit}
    \acro{ICAP}{Internal Configuration Access Port}
    \acro{IO}[I/O]{Input/Output}
    \acro{IP}{Intellectual Property}
    \acro{IV}{Initialization Vector}

    \acro{JTAG}{Joint Test Action Group}

    \acro{LUT}{Lookup Table}

    \acro{MAC}{Message Authentication Code}
    \acro{MATE}{Man-at-the-End}
    \acro{MITM}{Man-in-the-Middle}
    \acro{ML}{Machine Learning}

    \acro{NRE}{Non-Recurring Engineering}

    \acro{OTBN}{OpenTitan Big Number Accelerator}

    \acro{PC}{Program Counter}
    \acro{PCB}{Printed Circuit Board}
    \acro{PIP}{Programmable Interconnect Point}
    \acro{PKI}{Public Key Infrastructure}

    \acro{SRAM}{Static Random-Access Memory}
    \acro{RAM}{Random-Access Memory}
    \acro{RNG}{Random Number Generator}
    \acro{RST}{Remaining Silicon Thickness}
    \acro{RTL}{Register-Transfer Level}

    \acro{SEM}{Scanning Electron Microscope}
    \acro{STI}{Shallow Trench Isolation}
    \acro{SoC}{System-on-a-Chip}
    \acroplural{SoC}[SoCs]{Systems-on-Chip}
    
    \acro{TIFF}{Tag Image File Format}
    \acro{TLS}{Thermal Laser Stimulation}
    
    \acro{UART}{Universal Asynchronous Receiver Transmitter}
\end{acronym}

\begin{abstract}
\acp{FPGA} are known for their reprogrammability that allows for post-manufacture circuitry changes.
Nowadays, they are integral to a variety of systems including high-security applications such as aerospace and military systems. 
However, this reprogrammability also introduces significant security challenges, as bitstream manipulation can directly alter hardware circuits.
Malicious manipulations may lead to leakage of secret data and the implementation of hardware Trojans.
In this paper, we present a comprehensive framework for manipulating bitstreams with minimal reverse engineering, thereby exposing the potential risks associated with inadequate bitstream protection.
Our methodology does not require a complete understanding of proprietary bitstream formats or a fully reverse-engineered target design. 
Instead, it enables precise modifications by inserting pre-synthesized circuits into existing bitstreams. 
This novel approach is demonstrated through a semi-automated framework consisting of five steps: (1)~partial bitstream reverse engineering, (2)~designing the modification, (3)~placing and (4)~routing the modification into the existing circuit, and~(5) merging of the modification with the original bitstream.
We validate our framework through four practical case studies on the OpenTitan design synthesized for Xilinx 7-Series \acp{FPGA}.
While current protections such as bitstream authentication and encryption often fall short, our work highlights and discusses the urgency of developing effective countermeasures. 
We recommend using \acp{FPGA} as trust anchors only when bitstream manipulation attacks can be reliably excluded.
\end{abstract}




\maketitle

\section{Introduction}

Programmable hardware has become essential in today's digital landscape due to its adaptability, rapid development cycles, and low \acl{NRE} costs.
Consequently, \acfp{FPGA} are now pivotal components in safety-critical systems like medical devices, airplanes, network infrastructure, and defense applications.
These systems require an extremely high level of trust and reliability, making  hardware-based trust anchors indispensable.
Given their versatility, \acp{FPGA} are often employed as trust anchors in these critical applications~\cite{kataria2019}.
However, compared to \acp{ASIC}, the reconfigurability of the \ac{FPGA} fabric via bitstreams introduces additional security threats.
Though effective protection for bitstream authenticity, integrity, and confidentiality exist in theory, their implementations often have flaws that can leak information through side-channel attacks as well as flaws in the implementation itself~\cite{moradi2013,lohrke2018,Hettwer_Leger_Fennes_Gehrer_Gueneysu_2020, ender2020,rahman2021,CVEXilinxUSHack, skorobogatov2012breakthrough,ender2022,ICAPPaper}.

In this paper, we comprehensively investigate the threat potential of \ac{FPGA} design manipulations when their bitstream is not properly protected.
In the past, manipulation attacks were significantly hampered by the proprietary nature of bitstream formats.
In recent years, however, progress has been made in documenting widely used bitstream formats~\cite{ding2013,yu2018, benz2012,pham2017,bozzoli2018,ender2019,duncan2019, zhang2019,prjxray,kashani2022}, thus preparing the ground for reverse engineering and subsequent manipulation of deployed designs.
Yet it remains an extremely tedious and error-prone task to reverse engineer the bitstreams of complex real-world designs~\cite{klix2024}.
Thus, this paper posits that severe manipulations are possible with minimal reverse engineering of the underlying design.
Specifically, we propose and demonstrate a methodology, in which pre-synthesized circuit modifications are placed and routed on top of the original circuit, resulting in a functional bitstream containing both the original circuit and the intended manipulation.
Thereby, the manipulation circuit must first be designed at the \acf{RTL} and then placed and routed into unused fabric of the target chip.
\paragraph{Contribution}
We propose and develop a universal and efficient method to integrate complex manipulation circuits into existing bitstreams without having to reverse engineer the entire bitstream.
Our method requires only limited knowledge of the bitstream format, including the elements that make up the manipulation circuit, to ensure that the method does not inadvertedly override these elements in the targeted bitstream.
With this knowledge, we create a toolchain automating the processes of placing and routing synthesized \ac{FPGA} designs into an existing bitstream and connecting it to the already existing circuit.
We evaluate our proposed methodology through four comprehensive case studies that modify the OpenTitan design synthesized for a Xilinx' 7-Series \ac{FPGA}, highlighting its constructive and destructive potential.

\begin{figure}[H]
    \centering
    \includegraphics{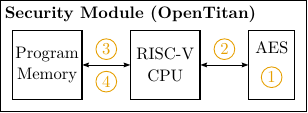} 
    \Description[The structure of the targeted attack potential within the chosen Security Module, the OpenTitan.]{The OpenTitan incorporates Program Memory, a RISC-V CPU, and an AES core. While case study 1 targets the AES module, case study 2 target the bus between the security module and the CPU, and case studies 3 and 4 target the data bus towards the program memory.}
    \caption{Manipulation entry points for our case studies.}
    \label{fpgamod:fig:casestudies}
\end{figure}

\paragraph{Case Studies}
Figure~\ref{fpgamod:fig:casestudies} shows the different components of the OpenTitan base design that are manipulated in our four case studies.

\begin{enumerate}[label=\protect\circledgold{\arabic*}]
    \item In the first case study, we demonstrate how to record and extract signal traces, \eg, from internal \ac{AES} states or for general debugging purposes.
    \item The second case study focuses on trojanizing the data flow while the secret key of an \ac{AES} instance is set.
    An attacker can utilize this to recover the secret key.
    \item The third case study demonstrates how to manipulate individual \ac{CPU} instructions as they flow between the program memory and the executing RISC-V \ac{CPU}, allowing an attacker to interfere with the initialization of the \ac{AES} unit.
    \item In the fourth case study, we replace an entire sequence of instructions to patch the running application and perform sophisticated attacks. 
    For demonstration purposes, we add a message block to the regular external communication interface containing the (protected) secret key.
\end{enumerate}

\paragraph{Availability.} We release all implementations required to reproduce the presented case studies under a permissive open-source license on GitHub: \url{https://github.com/emsec/PatchingFPGAs}

\section{Background \& Related Work}

\acp{FPGA} are reconfigurable, application-independent chips, whose operational circuitry is programmed by uploading a so-called bitstream.
Since we are using the widely deployed Xilinx 7-Series \acp{FPGA} in our case studies, we stick to their nomenclature.
The \ac{FPGA}'s reconfigurability is realized by a reprogrammable fabric consisting of programmable elements such as 6-to-2 \acp{LUT}, which contains the Boolean functions, \acp{FF}, and specialized blocks (\eg, \acp{DSP}). 
Several of these \acp{BEL} are organized in blocks and connected to each other by a reconfigurable routing realized by the so-called \acp{PIP}.

In the following, we situate our work within the existing literature and discuss the building blocks on which our work is based.

\subsection{Bitstream Protection Shortcomings}\label{fpgamod::background::bitstreamProtection}
To protect the users' designs and their \acl{IP} from manipulation, reverse engineering, and tampering, almost all \ac{FPGA} vendors have developed bitstream protection techniques such as bitstream encryption and authentication.
Although these techniques have been employed for decades, providing robust bitstream protection mechanisms has been a vexing problem for the \ac{FPGA} industry.
A large and constantly growing body of literature demonstrates that existing bitstream protection mechanisms can be defeated by side-channel and probing means~\cite{moradi2013,lohrke2018,moradi2016, Hettwer_Leger_Fennes_Gehrer_Gueneysu_2020}, as well as implementation flaws~\cite{ender2020,rahman2021,CVEXilinxUSHack, skorobogatov2012breakthrough,ender2022}.
Recently, Albartus~\etal~\cite{ICAPPaper} used the \ac{ICAP} to overcome bitstream protection schemes by exploiting this internal unsecured interface to design a stealthy Trojan implantation framework. 
In particular, they propose to hide hardware Trojans in the time and space domain by using an unsuspicious design that utilizes the \ac{ICAP} and then dynamically configures the Trojan.
In summary, such vulnerabilities and novel attack vectors demonstrate the de facto possibility of manipulating current \ac{FPGA} bitstreams, addressing this first challenge in bitstream manipulation.

\subsection{Bitstream Reverse Engineering}\label{fpgamod::background::bitstreamRE}
Another essential building block is the ability to understand the (proprietary) bitstream format.
Several works~\cite{ding2013,yu2018, benz2012,pham2017,bozzoli2018,ender2019,duncan2019} show the possibility of reverse engineering the bitstream format by creating a database that links the bits in the bitstream to the \acp{BEL} within the \ac{FPGA} fabric.
Additionally, various open-source resources document the bitstream format of different \acp{FPGA} and vendors~\cite{zhang2019,prjxray,kashani2022}, including most of the features of the Xilinx 7-Series.
While these tools are intended to generate bitstreams independently of the vendors' tools, we can use their databases to reverse engineer bitstreams to some extent.

\subsection{Bitstream Modifications Attacks}
Below, we review the implications and attack vectors from previous work on bitstream modification attacks.
For a comprehensive overview of recent bitstream modification attacks, see the work by Moraitis~\cite{DBLP:journals/access/Moraitis23}. 
In 2013, Chakraborty~\etal developed a method to merge two bitstreams of older generation Virtex-II \acp{FPGA}~\cite{chakraborty2013}.
In their case study, they added a circuit to dissipate power, resulting in faster aging of the device.
However, they work at the netlist level without using a reverse-engineered bitstream and do not mix the injected circuit with the existing circuit.

Swierczynski \etal~\cite{swierczynski2015} and Aldaya \etal~\cite{aldaya2016} randomly manipulated \acp{LUT} and \acp{BRAM} in order to fault \ac{FPGA} implementations of cryptographic algorithms, which would subsequently leak their secret keys.
Within this class of so-called bitstream fault injection attacks~\cite{swierczynski2017, engels2023, DBLP:conf/date/MoraitisD20, DBLP:conf/date/NiK0O24}, several other works inject modifications to fault a cryptographic computation within the design of the \ac{FPGA} to leak its secret key by means of classical fault injection attacks.
In summary, these works manipulate a single existing \ac{LUT} or net to fault the computation, rather than adding new components.

Also real-world products were attacked~\cite{swierczynski2016,kataria2019}. 
For example, Kataria \etal~\cite{kataria2019} defeated the trust anchor in Cisco Routers by manipulating the used bitstream. The advantage of their attack is that they have to reverse engineer and then manipulate the I/O pad configuration in the bitstream only.

The common feature of all previous work is that the changes made are minimal and mostly performed manually or with very little support from automated tools to ensure that the manipulations do not render the \ac{FPGA} design non-functional.
In contrast, our work interweaves modifications of different circuit sizes with the target designs and provides an automated toolchain that ensures that the target design is only affected by the developed modifications.
\section{Methodology \& Implementation}\label{fpgamod:sec:methodology}

In this section, we introduce the underlying threat model and describe our five-step methodology, along with implementation details for each step. % with which we integrate modifications into already placed and routed designs in bitstream format. %, further denoted as the original design.
\begin{figure*}[t]
    \centering
    \includegraphics[width=0.97\linewidth]{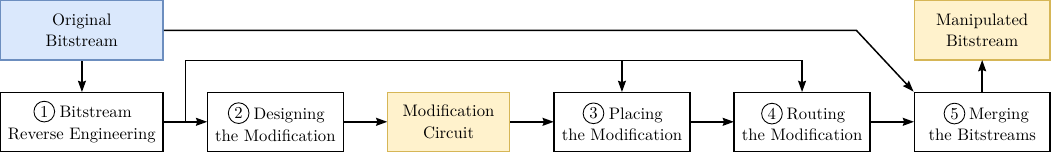}
    \Description[A row of 5 steps are shown. Input is the original bitstream and output is the manipulated bitstream.]{The first step is the bitstream reverse engineering, followed by the modification design, but also returning data required for placing the modification and routing the modification that follow after the modification circuit has been designed. The very last step is to merge the bitstreams.}
    \caption{Overview of the individual steps in our methodology.}
    \label{fpgamod:fig:overview}
\end{figure*}
As shown in Figure~\ref{fpgamod:fig:overview}, our general idea is to take an original bitstream, reverse engineer it and develop a modification circuit (in \ac{HDL}) to inject it into the original design by placing and routing it within the remaining resources in the \ac{FPGA}'s \acp{BEL}. % that is reverse engineered out of the original bitstream by use of a bitstream database like Project X-Ray~\cite{prjxray}.

\paragraph{Threat Model}
In our proposed threat, an attacker is assumed to have i)~access to the bitstream, ii)~the capability to interpret and manipulate it, and iii)~the ability to configure the manipulated bitstream back to the victim's \ac{FPGA}.
Access to the bitstream can be obtained by intercepting a (remote) update or firmware from a neighboring microcontroller, or by reading the bitstream directly from the non-volatile memory of an \ac{FPGA}.
If bitstream protection mechanisms are in place, our threat model assumes that these can be circumvented, as elaborated in Section~\ref{fpgamod::background::bitstreamRE}.
After gaining access, the attacker can reverse engineer the bitstream to locate manipulation points within the netlist and insert the malicious circuitry.
Finally, the attacker must program the manipulated bitstream back onto the target \ac{FPGA}.
This can be done using the same methods as for bitstream extraction, \eg, by uploading it to the \ac{FPGA} or non-volatile memory directly, or remotely using update mechanisms or adjacent microcontrollers.
Importantly, no \ac{HDL} code of the original design is required in our threat model, as all necessary design information is obtained via reverse engineering.

\subsection{Step\texorpdfstring{\,\circled{1}}{{ }1}: Partial Reverse Engineering of the Original Bitstream}
Once obtained, the original bitstream must be converted to a netlist.
The goal is to have a machine- and human-readable representation of the bitstream, which is used to find a suitable place to insert the manipulated design and to gain knowledge of where free \acp{BEL} are in the \ac{FPGA}.
The free resources are later used to place and route the modification design into the existing one.

\paragraph{Implementation.}
To achieve this goal for Xilinx 7-Series \acp{FPGA} we leveraged the open-source bitstream database of Project X-Ray~\cite{prjxray} and the accompanying fasm2bels tool.
This database was created using bitstream reverse engineering techniques as discussed in Section~\ref{fpgamod::background::bitstreamRE}.
The fasm2bels tool converts the bitstream into a \ac{FASM} file, which can be converted into a Verilog placed-and-routed netlist.
As some special cases and \acp{BEL} are missing in Project X-Ray's database, the resulting netlist is not entirely  complete.
However, this is not a problem since the most important \acp{BEL} are reversed, \ie, \acp{LUT}, \acp{FF}, and the routing, and in Step\,\circled{5} we will only add the modifications to the original bitstream instead of regenerating the whole bitstream.

\subsection{Step\texorpdfstring{\,\circled{2}}{{ }2}: Designing the Modification Circuit}
In the second step we define and implement the modification circuit. 
This step consists of the subtasks, first, to identify a suitable place to inject the modifications, and second, to implement the injected circuit in an \ac{HDL}.
Typically, a modification should interact with the original design, thus the analyst identifies the desired signals either by reverse engineering the design or by using high-level design information. 
The identified signals in the original design are then marked as \textit{listening} (\ie, input) or \textit{overriding} (\ie, output).
Then the analyst designs the modification circuit in an \ac{HDL} that uses the marked signals as its respective input and output ports.
Afterwards, the design is synthesized for the target \ac{FPGA} architecture.
This gives us a synthesized netlist that we can insert into the existing design in the following steps.

If a signal within a \ac{LUT}, \ie, an intermediate signal, is of interest, the corresponding \ac{LUT} must be decomposed.
This means that the Boolean function of the \ac{LUT} is split into two functions so that the intermediate signal can be tapped or manipulated.

\paragraph{Implementation.}
In our case studies, we implement several modification circuits ranging from implanting a logic analyzer to modifying the instruction register in OpenTitan.
Since we used the open-source OpenTitan project as the original design, we identified the target signals within the \ac{HDL} sources and mapped them to the placed and routed netlist.
If the analyst only has the bitstream, they can use established techniques and tools for netlist reverse engineering such as HAL and DANA~\cite{hal2019, albartus2020}.
We implemented all our modification circuits in SystemVerilog and synthesized them with Vivado.

\subsection{Step\texorpdfstring{\,\circled{3}}{{ }3}: Placing the Modification}

In this step, we start to merge the original circuit and the synthesized modification circuit by placing the elements from the modification circuit into the unused cells of the original design.
The routing is done in the next step.

\paragraph{Implementation.}
As part of the case studies, we implemented a custom placer in Python and Tcl that outputs an XDC file containing all the new placement information that can be used in Vivado.
It first replaces the input and output elements within the existing circuit to allow the signals connected to these cells to be replaced in the modification design.
All other cells are placed in resources not used by the original design.

To maintain timing closure, the cells are placed as close as possible to the connecting nets.
This is of significant interest, as each additional \ac{PIP} required for signal routing adds some delay to the critical path~\cite{ender2017}.
As we only use free tiles, we take the minimum feasible Euclidean distance between to be placed cells to other cells in the nets.
Special care has to be taken on placement of carry structure and muxes, as carry chains need to be aligned vertically in consecutive \acp{CLB} and in the correct bit order.
Muxes need to connect all \acp{LUT} signals that are input to the mux structure in two or four grouped \acp{LUT} of a \ac{CLB}.
Additionally, it reduces routing overhead when \acp{FF} are placed next their input, \ie, the \ac{LUT} feeding the \ac{FF}. 

\subsection{Step\texorpdfstring{\,\circled{4}}{{ }4}: Routing the Modification}

This step routes all of the added signals of the modified circuit, while respecting the existing routes of the original design by using only unused routing resources.
Hence, the timing of the original circuit is not changed to avoid timing closure issues.
The final result is a fully functional placed and routed netlist where the injected circuit is finally prepared for merging into the original design.

\paragraph{Implementation.}
We implemented a custom least-cost router in RapidWright~\cite{lavin2018} that meets the following requirements: i)~use only available unused routing resources and ii)~merge with the existing circuit.
For example, tapped input nets, \ie, existing signals that are fed into the modification circuit as inputs, can be branched at any junction of the existing route. 
Static net routing and clock routing must also be implemented, although the latter is ultimately so simple that it can be done by manually evaluating the clock tree and adding routing constraints to the modification design.
Whenever an unroutable situation occurs, our custom router prioritizes the unroutable net and repeats the routing process with the updated prioritized list of nets.

\subsection{Step\texorpdfstring{\,\circled{5}}{{ }5}: Merging the Bitstreams}

In the final step, the placed and routed modification circuit is converted into a bitstream which is then merged with the original bitstream.
Thus, only the bits in the original bitstream that are affected by the modification design are changed.
This way, elements that are not perfectly reverse engineered in Step\,\circled{1} will not be impaired.

\paragraph{Implementation.}
We create the final bitstream containing the original bitstream and the modification design as follows:
First,  the fully placed and routed modification circuit checkpoint file is loaded into Vivado to generate the bitstream. 
Only the modification circuit is converted into a bitstream, \ie, not the original circuit.
Second, the modification bitstream is converted to a \ac{FASM} file using the Project X-Ray database. 
Each line in the \ac{FASM} file represents a changed element compared to the original bitstream, \eg, a \ac{LUT} configuration or \ac{PIP} setting for the reprogrammable routing.
Third, the bitstreams are merged using a custom bitstream modification tool, that retrieves the bit positions of each manipulation specified in the \ac{FASM} file from the Project X-Ray database and modifies the corresponding bits in the original bitstream.

\section{Case Studies}
\label{fpgamod:sec:cs}

This section demonstrates our methodology through four case studies ranging from a passive recording of existing signals to the complete replacement of whole instruction sequences in the RISC-V \ac{CPU}, showing the adaptability and the possibilities for more sophisticated hardware Trojan implementations.
We used the OpenTitan, a silicon root of trust, for all case studies as our targeted FPGA design.
A required target application that runs on the OpenTitan is proposed in Section~\ref{fpgamod:sec:targetapp}.

\subsection{Target Application}
\label{fpgamod:sec:targetapp}
The OpenTitan includes various implementations of cryptographic primitives such as \ac{AES} and an Ibex RISC-V \ac{CPU} that runs the code that meets the needs of the respective security application and can access all peripheral modules.
One notable feature is its key manager, which is designed to securely store and transfer keys directly to cryptographic cores without involving the RISC-V core. % while being still under development at the time of our research, 
To perform cryptographic operations, a program must first be loaded into the RISC-V core to initialize registers and handle data.
Without such an application, the OpenTitan cannot perform any cryptographic operations.

\begin{algorithm}
 \begin{algorithmic}[1]
    \State $m\gets \text{UART}_{\text{RX}}()$
    \Loop
        \State $c\gets \text{enc}_k(m)$
        \State $\text{UART}_{\text{TX}}(c)$
        \State $m\gets m + 1$
    \EndLoop
 \end{algorithmic}
 \caption{\ac{AES} userspace application pseudocode.}
 \label{fpgamod:alg:app}
\end{algorithm}

Our target application for the OpenTitan continuously encrypts a message using the \ac{AES} core.
It initializes the OpenTitan, awaits a 16-byte initial plaintext over \ac{UART}, which is continuously encrypted and incremented by one while the resulting ciphertext is output over \ac{UART}.
Algorithm~\ref{fpgamod:alg:app} presents the pseudocode of this application (See Appendix~\ref{fpgamod:app:original} for test vectors).
This program mimics a typical cipher mode implementation, using multiple blocks to encrypt longer messages or to generate key streams.
Note that OpenTitan uses a two-share \ac{AES} hardware implementation, where we set one key share to a fixed key and the other to all zeros.
Using both key shares would give the same results and require both shares to be handled simultaneously, which is a minor methodological difference.

%%%%%%%%%%%%%%%%%%%%%%%%%%%%%%%%%%%%%%%%%%%%%%%%%%%%%%%%%%%%%%%%%%%%%%%
\subsection{Case Study 1: Extracting Signal Traces}
\label{fpgamod:sec:cs1}
%%%%%%%%%%%%%%%%%%%%%%%%%%%%%%%%%%%%%%%%%%%%%%%%%%%%%%%%%%%%%%%%%%%%%%%

Our first case study implements a logic analyzer that traces arbitrary signals present in the target circuit.
A similar bitstream-based approach was previously presented for older generation \acp{FPGA} by Graham~\cite{graham2001} and Hutchings~\cite{hutchings2014}, which we extend to the new 7-Series.
Such a logic analyzer can be used for (i)~dynamic analysis (as opposed to static analysis methods~\cite{azriel2019}), which may overcome hardware obfuscation by directly accessing internal signals, 
(ii)~leaking internal secret data such as key material,
and (iii)~serve as a debugging methodology during development without the need to resynthesize the entire design (as required by existing commercial and non-commercial logic analyzers~\cite{xilinxchipscope,sump2006,knittel2008}), thus saving time and possibly finding non-reproducible bugs in synthesizers since no resynthesis is required.
In our case study, we successfully connected the logic analyzer to the AES key register and leaked the used key.

\begin{figure}%[H]
    \centering
    \includegraphics[scale=0.68,trim=0.9cm 0 1.1cm 0]{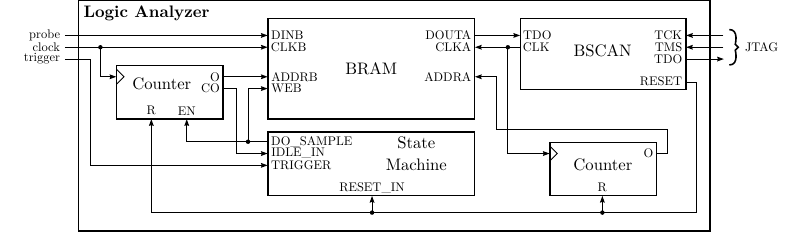}
    \Description[A block diagram showing the logic analyzer block detailed in 5 sub-blocks it consists of.]{The Write-Counter controls the BRAM memory address and the State Machine ensures that the memory block is filled with input signal traces. On the output side the read-address-counter is connected to the memory block as is the JTAG-interface namely the BSCAN module.}
    \caption{Case Study 1 Schematic (Extracting Signal Traces).}
    \label{fpgamod:fig:tinyla}
\end{figure}

The general structure of the tiny logic analyzer is shown in Figure~\ref{fpgamod:fig:tinyla}.
The logic analyzer stores the traced signals in a \ac{BRAM} and passes the recorded trace to the Boundary Scan (BSCAN) module, which is accessible through the \ac{JTAG} port and can be routed directly from the design.
Typically, the \ac{JTAG} interface is connected to at least one specific pin header on the \ac{PCB} for debugging purposes. % IEEE-SSCS-M paper selfcite?

A single trigger signal activates the sampling of up to 64 parallel probed signals by triggering a small state machine. 
The state machine enables writing to the \ac{BRAM} and its required address counter.
When either the \ac{BRAM} is full (2048 data points) or the trigger condition is false again, sampling is stopped and the traced content is dumped to the \ac{JTAG} port.
Notably, the logic analyzer does not affect the original circuit because the only output signal is routed to the newly instantiated BSCAN cell, and the overall routing of the original design remains the same, as intended by our methodology (\cf Section~\ref{fpgamod:sec:methodology}).

Using the computer software OpenOCD~\cite{openocd}, data is received from the \ac{JTAG} port.
The tool PulseView~\cite{sigrok} can display and decode received signal traces in an oscilloscope-like user interface.
In addition, the received data can be further analyzed to gain more insight into the target circuit.

%%%%%%%%%%%%%%%%%%%%%%%%%%%%%%%%%%%%%%%%%%%%%%%%%%%%%%%%%%%%%%%%%%%%%%%
\subsection{Case Study 2: Kleptographic Trojan}
\label{fpgamod:sec:cs2}
%%%%%%%%%%%%%%%%%%%%%%%%%%%%%%%%%%%%%%%%%%%%%%%%%%%%%%%%%%%%%%%%%%%%%%%

This case study leaks the secret key in OpenTitan's hardware \acs{AES}-128 in a kleptographic manner, a concept introduced by Young and Yung~\cite{young1996}.
We utilized the concept of weak kleptography by leaking the key through replacing a ciphertext output with the encrypted secret key, i.e., when the Trojan is activated, the ciphertext is replaced by $enc_{k_{Trojan}}(k)$.
By transmitting this ciphertext over an insecure channel, the attacker learns the original key by decrypting this kleptographic ciphertext, since they know $k_{Trojan}$.
If the same key is used for all encryptions, as in our target application (see Section~\ref{fpgamod:sec:targetapp}), the attacker can decrypt all further blocks.
However, the fact that the first block cannot be decrypted anymore with the original key is easily detectable.
Nevertheless, depending on the architecture of the implemented communication protocol, this might look similar to a communication problem and could lead to a message repetition that will be correctly encrypted.
Also, if the first block is relevant, \eg, for header information, the attacker is free to select a different block to replace with the key, only requiring a slightly more elaborate state machine or different trigger information.
An advantage of attacking the cryptographic module is that the proposed attack works whether the running program directly sets the key on the \ac{CPU} or the key originates from a secure key storage, like OpenTitan's key manager.
As a side note, this case study also applies to any other symmetric block cipher.

\begin{figure}%[H]
    \centering
    \includegraphics[scale=0.78,trim=1.7cm 0 1.7cm 0]{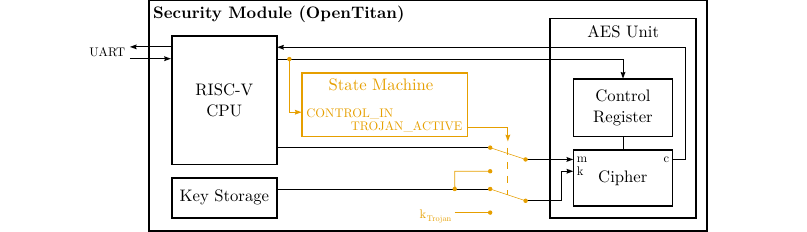}
    \Description[A block diagram showing the security module detailed in 4 relevant sub-blocks.]{The RISC-V CPU originally sends register information only to the AES unit. The Trojans state machine however taps into this bus and switches the connection from the key storage module to its own static key.}
    \caption{Case Study 2 Schematic (Kleptographic Trojan).}
    \label{fpgamod:fig:aestrojan}
\end{figure}

We have implemented this Trojan again in OpenTitan's \ac{AES} block, as shown in Figure~\ref{fpgamod:fig:aestrojan}.
Exemplary output is given in Appendix~\ref{fpgamod:app:cs2}.
The proposed Trojan consists of a small state machine for activation logic and muxes, switching between the legitimate input $k,m$ and our trojanized inputs $k_{Trojan},k$ to the \ac{AES} cipher block.
Since 128 key bits and 128 message bits are used, 256 muxes are inserted into the signals to the cipher block to decide between the trojanized input and the original input.

In order to perform the Trojan insertion into the circuit, the relevant key and state registers have to be found by reverse engineering means.
Prior work~\cite{albartus2020} has shown that finding 128-bit wide registers of an \ac{AES} implementation in the OpenTitan, including their correct bit order, is straightforward.
One method to find the correct register holding the key is to override each register one by one with a known value and then test if the encryption of the known values yields the expected value.
We assume the correct 128-bit key register and the 128-bit state register are correctly identified, as shown with previous methods.

%%%%%%%%%%%%%%%%%%%%%%%%%%%%%%%%%%%%%%%%%%%%%%%%%%%%%%%%%%%%%%%%%%%%%%%
\subsection{Case Study 3: Instruction Replacement Trojan}
\label{fpgamod:sec:cs3}
%%%%%%%%%%%%%%%%%%%%%%%%%%%%%%%%%%%%%%%%%%%%%%%%%%%%%%%%%%%%%%%%%%%%%%%

In this case study, we aim to integrate a hardware Trojan that modifies an instruction in the RISC-V \acs{CPU}.
For example, we change the loading of the encryption key in the running program, triggered by specific instructions.
In contrast to the previous case study 2, no manipulation of the cryptographic engine itself is required.
Thus, any eventual functional and timing-related checks in the cryptographic implementation will still pass.
Similarly, we assume that even when any code is signed or checked for modifications at runtime, the proposed hardware manipulation of the executed code will remain undetected, as the bus for instruction data accessed within the program code differs from the attacked one used for executing the instruction data.
Only replacing the key while keeping the payload the same is related to similar proposed hardware Trojans that weaken the cryptographic primitives~\cite{swierczynski2015}.
One application scenario where this typically remains undetected is cloud storage, where the same device encrypts and decrypts the data.
The encrypted blocks look like they are adequately encrypted but can nevertheless leak information to the attacker.

\begin{figure}%[H]
    \centering
    \includegraphics[scale=0.68,trim=0.9cm 0 0.9cm 0]{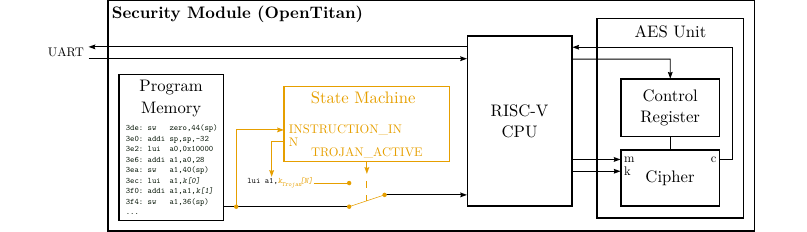}
    \Description[A block diagram showing the security module detailed in 4 relevant sub-blocks.]{The RISC-V CPU originally runs instructions coming from the program memory module and interacts with the AES module. The Trojans state machine however taps into the instruction bus and switches the data lines to replace parts of single instructions that load static key data.}
    \caption{Case Study 3 Schematic (Instruction Replacement Trojan).}
    \label{fpgamod:fig:replacement_trojan}
\end{figure}

A block diagram of the hardware Trojan circuit is shown in Figure~\ref{fpgamod:fig:replacement_trojan}.
Exemplary output is given in Appendix~\ref{fpgamod:app:cs3}.
When the key originates from the application program memory, intercepting the instruction bus gives an attacker complete control over the transferred key.
As the instruction and data buses are 32-bit wide each in the RISC-V, there will be subsequent instructions for loading a single 128-bit key.
The Trojan's logic is more complex and involves a counter for which word of the key to replace at which instruction.
The activation logic, however, only depends on the same instruction bus because the target application has a unique instruction that the Trojan can be triggered for just before loading the key material.

The instructions for loading the key into specified registers (load upper immediate and add immediate) can be patched directly to use a different key.
When the key is loaded at each block of encryption, depending on the application's implementation, the hardware Trojan can selectively affect single blocks of the encryption.
The manipulation's total effect is similar to the one in the previous case study.
The main difference is that the original key is not leaked to the attacker, but the attacker can decrypt the plaintext of selected blocks using the known key the hardware Trojan replaced the encryption key with.

%%%%%%%%%%%%%%%%%%%%%%%%%%%%%%%%%%%%%%%%%%%%%%%%%%%%%%%%%%%%%%%%%%%%%%%
\subsection{Case Study 4: Instruction Sequence Trojan}
\label{fpgamod:sec:cs4}
%%%%%%%%%%%%%%%%%%%%%%%%%%%%%%%%%%%%%%%%%%%%%%%%%%%%%%%%%%%%%%%%%%%%%%%

Our fourth case study extends the previous one by injecting sequences of instructions into a target application when a specific \ac{PC} address is reached in the running code.
Figure~\ref{fpgamod:fig:mod_trojan} shows a block diagram of the hardware Trojan circuit.
Exemplary output is given in Appendix~\ref{fpgamod:app:cs4}.
The replacement of instructions is again realized by a mux inserted between the instruction fetch and decode pipeline stages of the processor.
Once the Trojan is activated (state machine), the Trojan generates instructions with combinatorial logic (Trojan memory), as this approach provides a fast response whenever a new instruction is requested.
As our modification circuit listens to the address lines, the following instructions are overridden depending on the requested instruction address.

As an example, we have created a Trojan that inserts an additional ciphertext block containing $enc_{k_{Trojan}}(k)$, i.e., similar to the kleptographic Trojan (see Section~\ref{fpgamod:sec:cs2}), the original encryption key is encrypted with the key known to the attacker.
The additional message block is generated and sent when the application code sets the key.
This has the advantage that the secret key is already in a particular memory location and can easily be loaded as the plaintext to the \ac{AES} module together with the static attacker's key.
Then, the encryption is triggered, and the resulting ciphertext is sent over \ac{UART} before the original program execution continues.
In a well-defined communication protocol, this attack can be detected by the receiving party as an additional garbled message block that is received.
In this case, the key would need to be leaked out differently, probably not utilizing the \ac{UART}.
However, this is dependant on the Trojan use case and implementation.

\begin{figure}%[H]
    \centering
    \includegraphics[scale=0.68,trim=1cm 0 1cm 0]{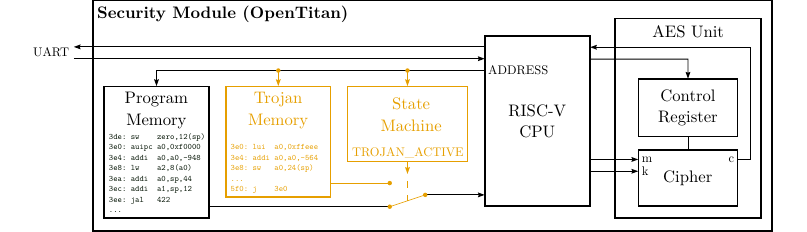}
    \Description[A block diagram showing the security module detailed in 5 relevant sub-blocks.]{The RISC-V CPU originally runs instructions coming from the program memory module and interacts with the AES module. The Trojans state machine however taps into the instruction bus and switches the data lines to its own trojan memory so larger parts of the program can be replaced, allowing to change security protocols.}
    \caption{Case Study 4 Schematic (Instruction Sequence Trojan).}
    \label{fpgamod:fig:mod_trojan}
\end{figure}

Finding a suitable location in the instruction data flow for Trojan insertion was complicated by tight timing constraints imposed by the critical path revealed at the running clock frequency.
As a note, we optimized our Trojan program to consist only of uncompressed RISC-V instructions, such that the critical path of the instruction decompressor is minimal.
The instruction decompressor, which handles compressed RISC-V instructions, is located in the same register stage as the one that our Trojan intercepts.

The Trojan program cannot be served as is to the \ac{CPU}.
It is required to handle the address logic of the original program, as the \acf{PC} is not affected by our Trojan instructions, except for jump instructions.
Instead, it is operating as if we would not inject our modification.
As RISC-V instructions can be either 2 or 4 bytes long, depending on whether they are compressed, the \ac{PC} can be incremented by 2 or 4 depending on the original instruction at the same address when executing an instruction.
This variance must be considered in the Trojan program by translating the addresses, especially in jump targets, and by relocating each instruction to the address present when the instruction is executed.

Finally, when the Trojan code has finished, it is desired to continue the original program, so the last Trojan instruction is to jump back to the initial trigger address.
The hardware Trojan's state logic disables the Trojan, and the original program continues to run where it was interrupted.

\section{Discussion}

This section summarizes the results and challenges of applying our methodology to the four case studies.
We discuss the outcomes, briefly review the pros and cons of open-source hardware, and suggest various countermeasures.

\subsection{Case Study Results}

\begin{figure*}[ht]
    \centering
    \begin{subfigure}[b]{0.22\linewidth}
        \centering
        \includegraphics[width=0.97\textwidth]{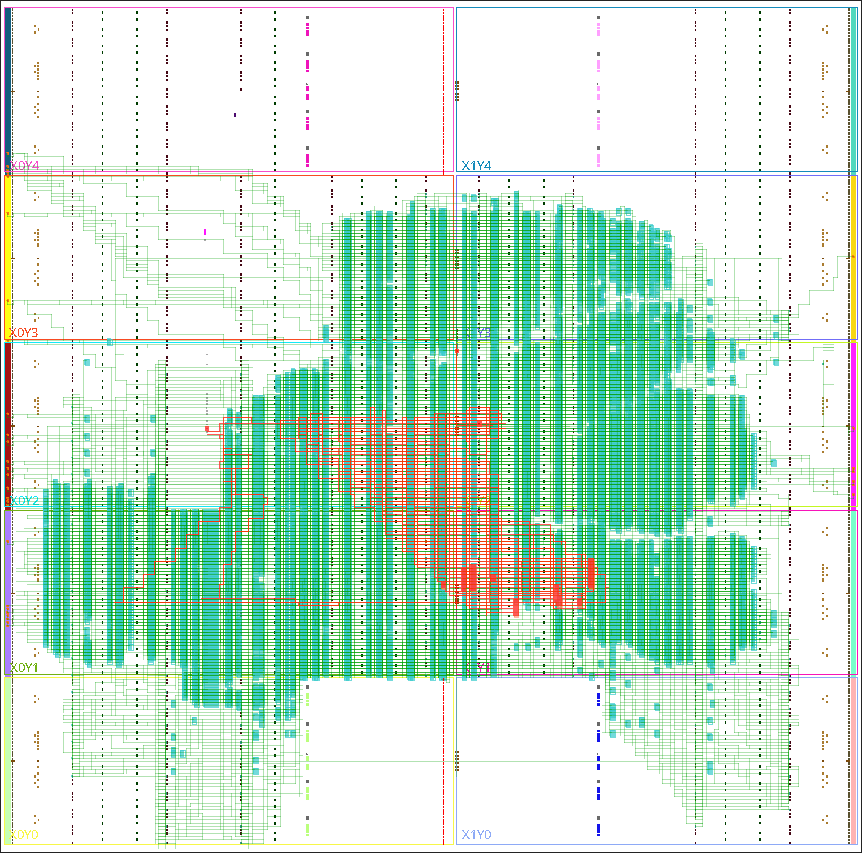}
        \caption{Extracting Signal Traces}
        \label{fpgamod:fig:device_tinyla}
    \end{subfigure}
    \hfill
    \begin{subfigure}[b]{0.22\linewidth}
        \centering
        \includegraphics[width=0.97\textwidth]{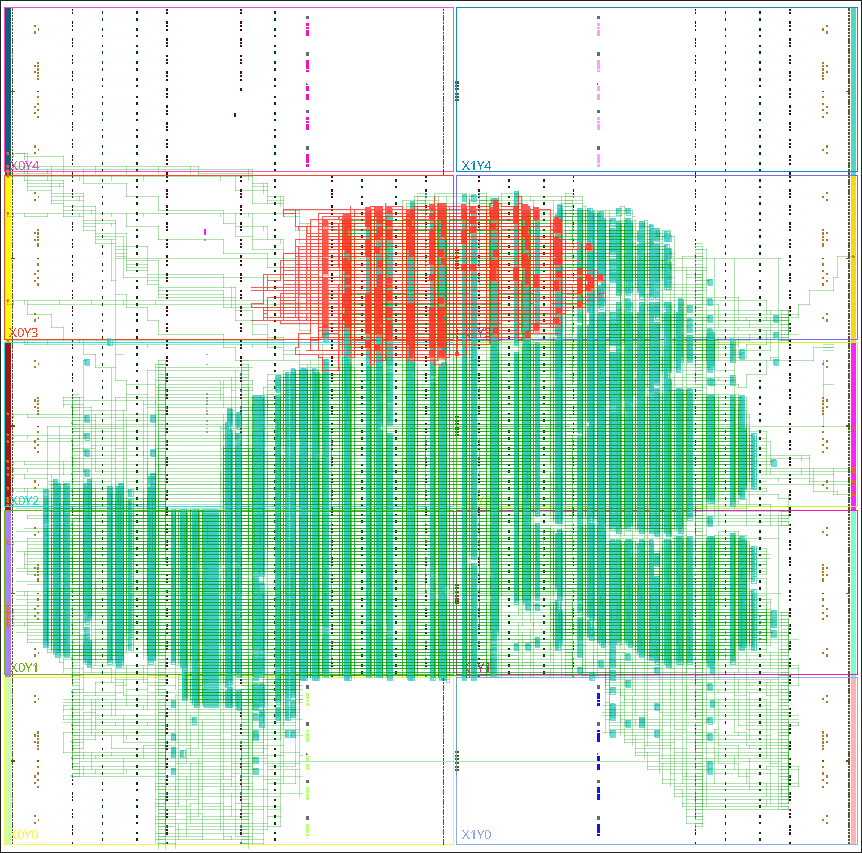}
        \caption{Kleptographic Trojan}
        \label{fpgamod:fig:device_aes}
    \end{subfigure}
    \hfill
    \begin{subfigure}[b]{0.22\linewidth}
        \centering
        \includegraphics[width=0.97\textwidth]{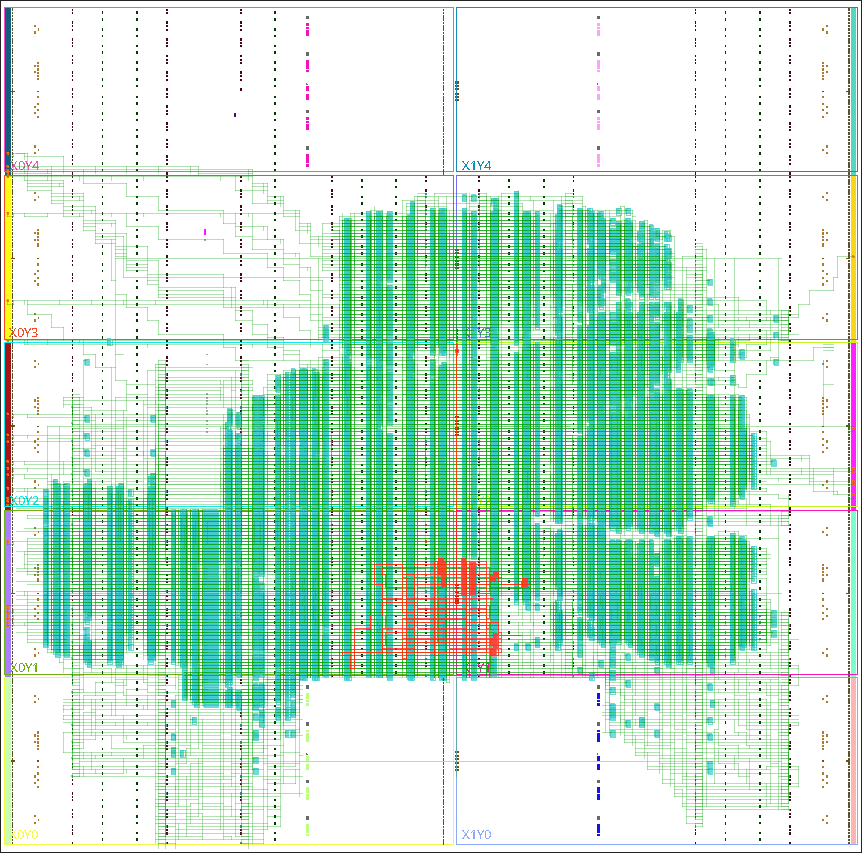}
        \caption{Instruction Replacement Trojan}
        \label{fpgamod:fig:device_instructionreplacement}
    \end{subfigure}
    \hfill
    \begin{subfigure}[b]{0.22\linewidth}
        \centering
        \includegraphics[width=0.97\textwidth]{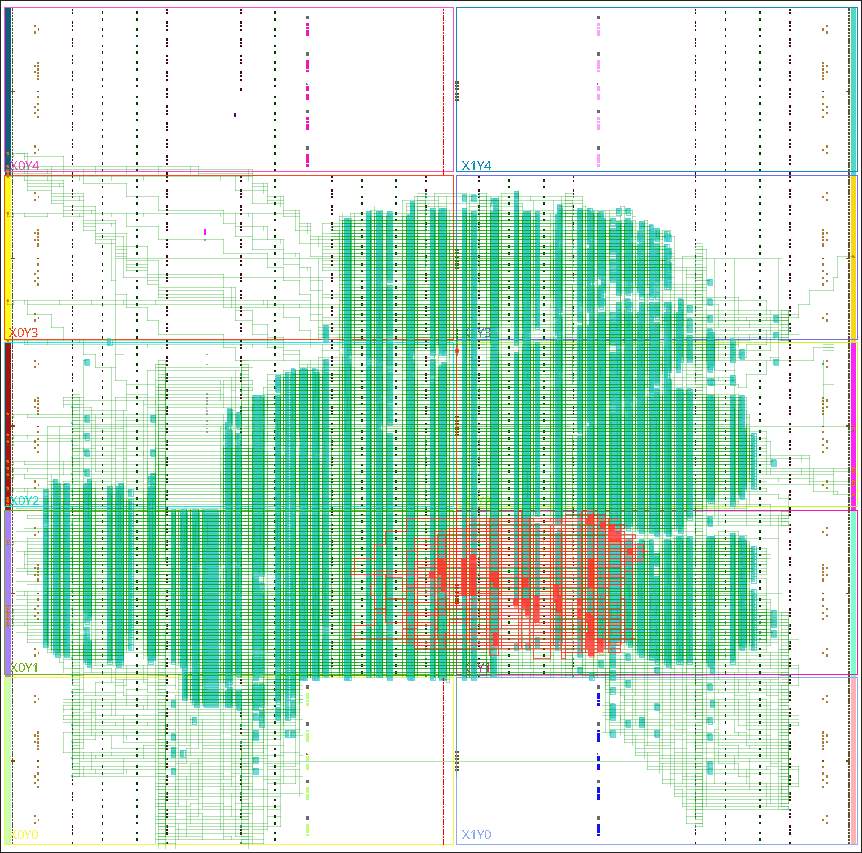}
        \caption{Instruction Sequence Trojan}
        \label{fpgamod:fig:device_instructionmod}
    \end{subfigure}
    \Description[4 subfigures showing the four utilizations of the single case studies on a device view of the Xilinx FPGA.]{In contrast, the original utilization of the OpenTitan circuit is taking up over half of the available space. The red highlighted case study modifications show that they are placed within areas of the original circuit. They are differently sized, while it seems that all but the third modification required more interconnections than the others. In general, the modification circuits only affect small parts of the whole fabric.}
    \caption{Device utilization view of the \ac{FPGA}. The original OpenTitan design is shown in green, while the modified circuit of each case study is shown in red.}
    \label{fpgamod:fig:device}
\end{figure*}

We successfully implemented our four comprehensive case studies.
For each case study, we created a modification to the original OpenTitan design to be placed and routed based on the information we obtained from the original bitstream.
Figure~\ref{fpgamod:fig:device} shows the \ac{FPGA} device utilization with the OpenTitan design and the overlaid modifications.
The OpenTitan continued to function with these changes, which performed as expected when triggered by the designed activation events. 
For testing, we used a Xilinx Artix-7 200T on a Digilent Nexys Video development board. 
Our results are reproducible on any 7-Series \ac{FPGA}, with minor variations due to resource availability and routing differences.

We present the relevant statistics of our results in Table~\ref{fpgamod:tab:details}.
The first case study does not override any signals in the target design.
Thus, the original design always functions the same.
The only inputs to the tiny logic analyzer are the 64 signals for recording the traces.
The low logic overhead of the first case study results in low placement and routing times.
Automated injection of the logic analyzer circuit into the OpenTitan takes less than 2 minutes, not including the time needed for development of the modification design and for intermediate synthesis processes of Vivado.
For example, once a fixed design like the tiny logic analyzer is implemented, it can be integrated into a random design using automated methods more quickly than re-synthesizing larger complete \ac{FPGA} designs.
This fast processing time is advantageous for debugging arbitrary signals during the design phase.
Furthermore, placement and routing can likely be optimized beyond our unoptimized implementation, which ran on a modern laptop with an Intel Core i7-8665U processor and 40 GB of \ac{RAM}.

\begin{table}[ht]
    \centering
    \caption{Statistics of the implemented case studies. For comparison, the full Open Titan base design consists of 51756 \acp{LUT}, 22839 \acp{FF}, 2401 \ac{RAM} blocks, 1307 CARRY4, and 1742 MUXFX instances.
}
    \label{fpgamod:tab:details}
		\begin{tabular}{l|rrrr}
			\toprule
			& \textbf{CS 1} & \textbf{CS 2} & \textbf{CS 3} & \textbf{CS 4} \\
			\midrule
            Signals Sniffed (Inputs) & 64 & 261 & 27 & 66 \\
            Signals Overridden (Outputs) & 0 & 256 & 20 & 38 \\
            Net Pins Overridden & 0 & 384 & 208 & 399 \\[0.2cm]
            Additional \acsp{LUT} & 34 & 257 & 36 & 412 \\
            \acsp{LUT} Decomposited & 0 & 0 & 0 & 4 \\
            Additional \acsp{FF} & 27 & 1 & 6 & 2 \\
            Additional \acsp{BRAM} & 1 & 0 & 0 & 0 \\
            Additional BSCANs & 1 & 0 & 0 & 0 \\
            Additional CARRY4s & 4 & 0 & 0 & 0 \\
            Additional MUXFXs & 0 & 0 & 0 & 46 \\[0.2cm]
            Total Nets & 146 & 519 & 69 & 530 \\
            Total \acsp{PIP} & 2888 & 6223 & 1431 & 9232 \\
            Longest Route in \acsp{PIP} & 73 & 52 & 42 & 51 \\[0.2cm]
            Bits Modified (0 $\rightarrow$ 1) & 8541 & 21756 & 3803 & 30958 \\
            Bits Modified (1 $\rightarrow$ 0) & 0 & 1845 & 897 & 1757 \\[0.2cm]
            Time \circled{3} Placement & 10\,s & 114\,s & 10\,s & 296\,s \\
            Time \circled{4} Routing  & 48\,s & 101\,s & 33\,s & 161\,s \\
            Time \circled{5} Merging Bitstreams  & 50\,s & 50\,s & 40\,s & 45\,s \\
            Time Total & 108\,s & 265\,s & 83\,s & 502\,s \\
			\bottomrule
		\end{tabular}
\end{table}

The second case study requires many input and output signals, as the 128-bit key and state registers of the \ac{AES} module must be intercepted.
The numerous input signals increase the number of nets to be routed, leading to longer routing times and more than doubling the differing bits between the modified and original bitstreams compared to the first case study.
The automated steps of the cryptographic implementation Trojan injection take less than 5 minutes.

The third case study requires the fewest signal inputs and outputs and has the smallest Trojan logic.
This results in fewer differing bits between the bitstreams.
The automated steps were executed in 83 seconds.
Compared to the half-hour design run of the OpenTitan, this is a significant improvement over the naive strategy of reverse engineering the original bitstream, modifying the circuit, and re-synthesizing the whole design.
Larger designs for bigger \acp{FPGA} can take several hours to synthesize, place, and route, but our technique introduces modifications in a consistently short time, largely independent of the target design's size.

The fourth case study requires fewer inputs and outputs than the second but has larger internal logic due to the fully implemented combinatorial logic of the replaced instruction sequence.
The number of logic elements relates to the size of the program modification, resulting in the most changed bits in the bitstream—4 KiB out of 9.3 MiB.
Despite the extensive modifications, the runtime is still below 10 minutes.

Routing complexity can be estimated by the relation between the total number of \acp{PIP} and the number of nets to be routed. 
A high difference between overridden net pins and actual overridden output signals indicates a large fan-out, increasing routing interactions with the original design and utilizing more \acp{PIP}. 
The longest route should be considered for signal delay, as timing requirements vary. 

\subsection{Implementation Challenges}

Challenges in implementation arose from various factors. 
Most issues related to unexpected behavior after synthesis and injection into the target bitstream. 
Insufficient routing capabilities, incorrect placement constraints, and previously occupied cells or pins were common problems. 
In such cases, the OpenTitan would either not respond to \ac{UART} commands or enter a fail state when the Trojan was triggered, sometimes outputting an error message or just halting the \ac{CPU}.
Specific issues during automated routing included:
\begin{itemize}
\item Not considering both directions for bidirectional \acp{PIP}. Sometimes, such a \ac{PIP} was occupied in the original design, so the other direction cannot be used by the modification.
\item Some \acp{PIP} in \ac{CLB} interfaces are routed to alternative pins to improve routing efficiency. However, these alternative pins may have a secondary connection always routed into the \ac{CLB}, which must be considered if specific \ac{CLB} features are in use.
\item Database mismatches between Vivado and RapidWright causing unroutable situations or invalid database values.\footnote{Specifically, the relation between so-called \textit{RouteNodes} and their respective \textit{Wires}. We reported these issues to Xilinx to improve future releases of RapidWright.}
\end{itemize}

The automated tools worked effectively once these issues were addressed.
These problems mainly occurred due to inaccessible knowledge about and the complexity of the target architecture. 
Prior research on bitstream formats and reverse engineering efforts, such as Project x-Ray~\cite{prjxray}, proved invaluable in overcoming the issues. 
Workarounds included avoiding global \ac{IO} pins and using completely vacant slices for modifications.

During the modification design phase (Step\,\texorpdfstring{\circled{2}}{2}), simulating outputs on defined input traces, including the circuit trigger, was beneficial. 
The open-source nature of OpenTitan allowed us to integrate its functionality into the \ac{RTL} code for testing before injection. 
In scenarios without access to the original design, simulation tools in reverse engineering toolkits like HAL~\cite{hal2019} and Verilator~\cite{verilator2007} can help simulate behavior on reverse-engineered netlist parts. 
Testing small parts of the manipulation in isolation simplified issue identification.

Signal timing had minimal impact on modification functionality, as long as signal delay did not exceed the clock tick duration. 
While timing validation methods are advisable for larger modifications or more compact designs, our modifications did not require such measures. 
Keeping routing short reduces timing load, and re-ordering logic cells or rerouting logic layers can resolve timing issues. 
Alternatively, injecting similar manipulations elsewhere can also address these challenges.

\subsection{Open Source}
From an attacker's perspective, designing hardware Trojans is easier when the \ac{RTL} code of the design under attack is available, as with OpenTitan.
While closed sources increase the effort to craft meaningful attacks based solely on the bitstream, they cannot prevent them entirely.
According to Kerkhoff's principle, a system's security should not rely on keeping its construction secret.
Open-source designs can be reviewed and subsequently hardened by the community, making attacks more challenging, whereas closed-source systems may harbor exploitable flaws, benefiting attackers.

\subsection{Countermeasures}
This section explores countermeasures to safeguard \ac{FPGA} designs against undesired bitstream manipulations. 
Effective solutions often necessitate comprehensive protection measures or chip-level modifications, precluding upgrades to enhance security in existing \acp{FPGA}. 
A layered approach that combines multiple countermeasures appears essential for effectively mitigating bitstream manipulation attacks.

\begin{description}
\item[Physical Access Protection]
Protecting the bitstream from physical extraction and replacement is critical when other technological protections fail.
Preventing access to both \ac{SRAM}-based \ac{FPGA} and persistent memory (\eg, flash) and securing the data bus between them is essential. 
Advanced \acp{HSM} include features like backup batteries for data deletion upon intrusion detection, mesh traces for tamper detection, and novel methods such as radio wave response verification~\cite{staat2022} and physical movement detection~\cite{goette2021} to prevent unauthorized access. 
However, these methods often come with significant implementation overhead, making them less practical for widespread application.

\item[Bitstream Security]
Encrypting and authenticating bitstreams is a standard method to prevent unauthorized reverse engineering and modifications.
Hence, such schemes would directly hinder attacks like those shown in our work.
However, as discussed in Section~\ref{fpgamod::background::bitstreamProtection}, most bitstream protection schemes are vulnerable to either side-channel attacks or implementation flaws.

\item[Obfuscation Techniques]
Obfuscation of bitstream contents adds a layer of protection against reverse engineering of a design's netlist. 
These techniques often aim to obscure the functional details of \acp{LUT}~\cite{karam2016,hazari2018,kamali2018,olney2020}, making it challenging for attackers to understand the underlying design without first breaking the obfuscation. 
While obfuscation increases the complexity for analysts or attackers attempting to interpret the netlist, it does not prevent malicious manipulations.

\item[Filling \ac{FPGA} Fabric]
\ac{FPGA} designs often do not fully utilize available fabric resources. 
By introducing elaborate dummy signals and populating redundant or unused logic cells throughout the \ac{FPGA}, resource saturation can be increased.
This countermeasure aims to thwart methodologies that seek successful placements and routings for modification circuits, potentially imposing tighter timing constraints. 
However, efficiently generating circuits that continue to function seamlessly remains an ongoing challenge with this approach.

\item[Self-Tests]
A common countermeasure is self-testing, either within the circuit or within the CPU, to test pre-calculated values or control flows.
For example, one countermeasure implemented in OpenTitan is the integration of alert mechanisms into the hardware to detect abnormal states.
Similarly, \ac{CPU} feature checks implemented in \acp{FPGA} could be performed on the running program, \eg, control flow checks, and similar measures to protect against running unauthorized software.
However, such alerting mechanisms are ineffective against internal manipulation attacks, as these attacks may not trigger alerts or may be defeated by turning them off.
Nevertheless, this countermeasure makes it more difficult for an attacker, but does not prevent attacks.

\item[Alternative Key Storage and Secure Cores]
The OpenTitan features more secure ways to handle security-critical operations than executing \ac{CPU} instructions, namely the key manager and the \ac{OTBN} core.
With these, any key cannot be leaked over the instruction bus as it is covered in secure sub-modules of the OpenTitan.
However, the respective data lines handling the key inside these modules might be tapped with a logic analyzer or tampered with otherwise.
Furthermore, it would be possible to introduce attacks on the instruction bus using securely implemented algorithms.
This attack could, for instance, lead to an oracle by executing cryptographic operations on the now-secured key.
The challenge to find the right spot by reverse engineering to introduce the attack remains the same.
The \ac{AES} core Trojan developed in Section~\ref{fpgamod:sec:cs2} does not protect against usage of the key manager as the key origin is irrelevant for this attack, so it works the same even when the key is not loaded directly by the user space program.

\end{description}
\section{Conclusion}

This paper demonstrates the feasibility of inserting sophisticated hardware Trojans into \ac{FPGA} bitstreams with minimal reverse engineering efforts. 
Our automated toolchain efficiently integrates pre-synthesized circuit modifications into existing designs.
The four successful case studies targeting an OpenTitan \ac{FPGA} implementation range from extracting signal traces from an \ac{AES} core, over obtaining secret keys over the \ac{UART} interface, to replacing secret keys. 

Our findings underscore the vulnerability of \ac{FPGA} designs when bitstreams are inadequately protected. 
While our method requires basic knowledge of the bitstream format, it exposes the potential for malicious modifications using limited resources.

Future research should further evaluate countermeasures to protect \ac{FPGA} designs against unwanted bitstream manipulations discussed in this work and explore the feasibility of adapting our approach to \acp{ASIC}, which requires to perform more elaborated edits, \ie, using a \acl{FIB} or by directly targeting the chip design or production files.
Continued efforts in standardizing and securing open-source hardware will be crucial for balancing transparency with robust protection against hardware-level attacks.

\begin{acks}
We thank Julian Speith for his efforts put in the development of the bitstream merging tool.
This work was supported in part by the Deutsche Forschungsgemeinschaft (DFG, German Research Foundation) under Germany's Excellence Strategy -- EXC 2092 CASA –- 390781972 and by the Research Center Trustworthy Data Science and Security (\url{https://rc-trust.ai}), one of the Research Alliance Centers within the UA Ruhr (\url{https://uaruhr.de}).
\end{acks}

\bibliographystyle{ACM-Reference-Format}
\bibliography{biblio}

%%% -*-BibTeX-*-
%%% Do NOT edit. File created by BibTeX with style
%%% ACM-Reference-Format-Journals [18-Jan-2012].

\begin{thebibliography}{51}

%%% ====================================================================
%%% NOTE TO THE USER: you can override these defaults by providing
%%% customized versions of any of these macros before the \bibliography
%%% command.  Each of them MUST provide its own final punctuation,
%%% except for \shownote{}, \showDOI{}, and \showURL{}.  The latter two
%%% do not use final punctuation, in order to avoid confusing it with
%%% the Web address.
%%%
%%% To suppress output of a particular field, define its macro to expand
%%% to an empty string, or better, \unskip, like this:
%%%
%%% \newcommand{\showDOI}[1]{\unskip}   % LaTeX syntax
%%%
%%% \def \showDOI #1{\unskip}           % plain TeX syntax
%%%
%%% ====================================================================

\ifx \showCODEN    \undefined \def \showCODEN     #1{\unskip}     \fi
\ifx \showDOI      \undefined \def \showDOI       #1{#1}\fi
\ifx \showISBNx    \undefined \def \showISBNx     #1{\unskip}     \fi
\ifx \showISBNxiii \undefined \def \showISBNxiii  #1{\unskip}     \fi
\ifx \showISSN     \undefined \def \showISSN      #1{\unskip}     \fi
\ifx \showLCCN     \undefined \def \showLCCN      #1{\unskip}     \fi
\ifx \shownote     \undefined \def \shownote      #1{#1}          \fi
\ifx \showarticletitle \undefined \def \showarticletitle #1{#1}   \fi
\ifx \showURL      \undefined \def \showURL       {\relax}        \fi
% The following commands are used for tagged output and should be
% invisible to TeX
\providecommand\bibfield[2]{#2}
\providecommand\bibinfo[2]{#2}
\providecommand\natexlab[1]{#1}
\providecommand\showeprint[2][]{arXiv:#2}

\bibitem[Albartus et~al\mbox{.}(2024)]%
        {ICAPPaper}
\bibfield{author}{\bibinfo{person}{Nils Albartus}, \bibinfo{person}{Maik
  Ender}, \bibinfo{person}{Jan{-}Niklas M{\"{o}}ller}, \bibinfo{person}{Marc
  Fyrbiak}, \bibinfo{person}{Christof Paar}, {and} \bibinfo{person}{Russell
  Tessier}.} \bibinfo{year}{2024}\natexlab{}.
\newblock \showarticletitle{On the Malicious Potential of Xilinx's Internal
  Configuration Access Port {(ICAP)}}.
\newblock \bibinfo{journal}{\emph{{ACM} Trans. Reconfigurable Technol. Syst.}}
  \bibinfo{volume}{17}, \bibinfo{number}{2} (\bibinfo{year}{2024}),
  \bibinfo{pages}{26:1--26:28}.
\newblock
\urldef\tempurl%
\url{https://doi.org/10.1145/3633204}
\showDOI{\tempurl}


\bibitem[Albartus et~al\mbox{.}(2020)]%
        {albartus2020}
\bibfield{author}{\bibinfo{person}{Nils Albartus}, \bibinfo{person}{Max
  Hoffmann}, \bibinfo{person}{Sebastian Temme}, \bibinfo{person}{Leonid
  Azriel}, {and} \bibinfo{person}{Christof Paar}.}
  \bibinfo{year}{2020}\natexlab{}.
\newblock \showarticletitle{DANA Universal Dataflow Analysis for Gate-Level
  Netlist Reverse Engineering}.
\newblock \bibinfo{journal}{\emph{IACR Transactions on Cryptographic Hardware
  and Embedded Systems}} \bibinfo{volume}{2020}, \bibinfo{number}{4}
  (\bibinfo{date}{Aug.} \bibinfo{year}{2020}), \bibinfo{pages}{309–336}.
\newblock
\urldef\tempurl%
\url{https://doi.org/10.13154/tches.v2020.i4.309-336}
\showDOI{\tempurl}


\bibitem[Azriel et~al\mbox{.}(2019)]%
        {azriel2019}
\bibfield{author}{\bibinfo{person}{Leonid Azriel}, \bibinfo{person}{Ran
  Ginosar}, {and} \bibinfo{person}{Avi Mendelson}.}
  \bibinfo{year}{2019}\natexlab{}.
\newblock \showarticletitle{SoK: An Overview of Algorithmic Methods in IC
  Reverse Engineering}. In \bibinfo{booktitle}{\emph{Proceedings of the 3rd ACM
  Workshop on Attacks and Solutions in Hardware Security Workshop}} (London,
  United Kingdom). \bibinfo{publisher}{Association for Computing Machinery},
  \bibinfo{address}{New York, NY, USA}, \bibinfo{pages}{65–74}.
\newblock
\showISBNx{9781450368391}
\urldef\tempurl%
\url{https://doi.org/10.1145/3338508.3359575}
\showDOI{\tempurl}


\bibitem[Benz et~al\mbox{.}(2012)]%
        {benz2012}
\bibfield{author}{\bibinfo{person}{Florian Benz}, \bibinfo{person}{André
  Seffrin}, {and} \bibinfo{person}{Sorin~A. Huss}.}
  \bibinfo{year}{2012}\natexlab{}.
\newblock \showarticletitle{Bil: A tool-chain for bitstream
  reverse-engineering}. In \bibinfo{booktitle}{\emph{22nd International
  Conference on Field Programmable Logic and Applications (FPL)}}.
  \bibinfo{pages}{735--738}.
\newblock
\urldef\tempurl%
\url{https://doi.org/10.1109/FPL.2012.6339165}
\showDOI{\tempurl}


\bibitem[Bozzoli and Sterpone(2018)]%
        {bozzoli2018}
\bibfield{author}{\bibinfo{person}{Ludovica Bozzoli} {and}
  \bibinfo{person}{Luca Sterpone}.} \bibinfo{year}{2018}\natexlab{}.
\newblock \showarticletitle{COMET: a Configuration Memory Tool to Analyze,
  Visualize and Manipulate FPGAs Bitstream}. In \bibinfo{booktitle}{\emph{ARCS
  Workshop 2018; 31th International Conference on Architecture of Computing
  Systems}}. \bibinfo{pages}{1--4}.
\newblock


\bibitem[Cabrera~Aldaya et~al\mbox{.}(2016)]%
        {aldaya2016}
\bibfield{author}{\bibinfo{person}{Alejandro Cabrera~Aldaya},
  \bibinfo{person}{Alejandro Cabrera-Sarmiento}, {and}
  \bibinfo{person}{Santiago Sánchez-Solano}.} \bibinfo{year}{2016}\natexlab{}.
\newblock \showarticletitle{AES T-Box tampering attack}.
\newblock \bibinfo{journal}{\emph{Journal of Cryptographic Engineering}}
  \bibinfo{volume}{6} (\bibinfo{date}{04} \bibinfo{year}{2016}),
  \bibinfo{pages}{31--48}.
\newblock
\urldef\tempurl%
\url{https://doi.org/10.1007/s13389-015-0103-4}
\showDOI{\tempurl}


\bibitem[Chakraborty et~al\mbox{.}(2013)]%
        {chakraborty2013}
\bibfield{author}{\bibinfo{person}{Rajat~Subhra Chakraborty},
  \bibinfo{person}{Indrasish Saha}, \bibinfo{person}{Ayan Palchaudhuri}, {and}
  \bibinfo{person}{Gowtham~Kumar Naik}.} \bibinfo{year}{2013}\natexlab{}.
\newblock \showarticletitle{Hardware Trojan Insertion by Direct Modification of
  FPGA Configuration Bitstream}.
\newblock \bibinfo{journal}{\emph{IEEE Design \& Test}} \bibinfo{volume}{30},
  \bibinfo{number}{2} (\bibinfo{year}{2013}), \bibinfo{pages}{45--54}.
\newblock
\urldef\tempurl%
\url{https://doi.org/10.1109/MDT.2013.2247460}
\showDOI{\tempurl}


\bibitem[Dang~Pham et~al\mbox{.}(2017)]%
        {pham2017}
\bibfield{author}{\bibinfo{person}{Khoa Dang~Pham}, \bibinfo{person}{Edson
  Horta}, {and} \bibinfo{person}{Dirk Koch}.} \bibinfo{year}{2017}\natexlab{}.
\newblock \showarticletitle{BITMAN: A tool and API for FPGA bitstream
  manipulations}. In \bibinfo{booktitle}{\emph{Design, Automation \& Test in
  Europe Conference \& Exhibition (DATE), 2017}}. \bibinfo{pages}{894--897}.
\newblock
\urldef\tempurl%
\url{https://doi.org/10.23919/DATE.2017.7927114}
\showDOI{\tempurl}


\bibitem[Ding et~al\mbox{.}(2013)]%
        {ding2013}
\bibfield{author}{\bibinfo{person}{Zheng Ding}, \bibinfo{person}{Qiang Wu},
  \bibinfo{person}{Yizhong Zhang}, {and} \bibinfo{person}{Linjie Zhu}.}
  \bibinfo{year}{2013}\natexlab{}.
\newblock \showarticletitle{Deriving an NCD file from an FPGA bitstream:
  Methodology, architecture and evaluation}.
\newblock \bibinfo{journal}{\emph{Microprocess. Microsystems}}
  \bibinfo{volume}{37} (\bibinfo{year}{2013}), \bibinfo{pages}{299--312}.
\newblock
\urldef\tempurl%
\url{https://api.semanticscholar.org/CorpusID:39426622}
\showURL{%
\tempurl}


\bibitem[Duncan et~al\mbox{.}(2019)]%
        {duncan2019}
\bibfield{author}{\bibinfo{person}{Adam Duncan}, \bibinfo{person}{Fahim
  Rahman}, \bibinfo{person}{Andrew Lukefahr}, \bibinfo{person}{Farimah
  Farahmandi}, {and} \bibinfo{person}{Mark~Mohammad Tehranipoor}.}
  \bibinfo{year}{2019}\natexlab{}.
\newblock \showarticletitle{FPGA Bitstream Security: A Day in the Life}.
\newblock \bibinfo{journal}{\emph{2019 IEEE International Test Conference
  (ITC)}} (\bibinfo{year}{2019}), \bibinfo{pages}{1--10}.
\newblock
\urldef\tempurl%
\url{https://api.semanticscholar.org/CorpusID:211227213}
\showURL{%
\tempurl}


\bibitem[{Embedded Security Group}(2019)]%
        {hal2019}
\bibfield{author}{\bibinfo{person}{{Embedded Security Group}}.}
  \bibinfo{year}{2019}\natexlab{}.
\newblock \bibinfo{title}{{HAL - The Hardware Analyzer}}.
\newblock
\newblock
\urldef\tempurl%
\url{https://github.com/emsec/hal}
\showURL{%
Retrieved July 4, 2024 from \tempurl}


\bibitem[Ender et~al\mbox{.}(2017)]%
        {ender2017}
\bibfield{author}{\bibinfo{person}{Maik Ender}, \bibinfo{person}{Samaneh
  Ghandali}, \bibinfo{person}{Amir Moradi}, {and} \bibinfo{person}{Christof
  Paar}.} \bibinfo{year}{2017}\natexlab{}.
\newblock \showarticletitle{The First Thorough Side-Channel Hardware Trojan}.
  In \bibinfo{booktitle}{\emph{ASIACRYPT (1)}}. \bibinfo{publisher}{Springer},
  \bibinfo{pages}{755--780}.
\newblock
\urldef\tempurl%
\url{https://doi.org/10.1007/978-3-319-70694-8_26}
\showDOI{\tempurl}


\bibitem[Ender et~al\mbox{.}(2022)]%
        {ender2022}
\bibfield{author}{\bibinfo{person}{Maik Ender}, \bibinfo{person}{Gregor
  Leander}, \bibinfo{person}{Amir Moradi}, {and} \bibinfo{person}{Christof
  Paar}.} \bibinfo{year}{2022}\natexlab{}.
\newblock \showarticletitle{A Cautionary Note on Protecting Xilinx’
  UltraScale(+) Bitstream Encryption and Authentication Engine}. In
  \bibinfo{booktitle}{\emph{2022 IEEE 30th Annual International Symposium on
  Field-Programmable Custom Computing Machines (FCCM)}}. \bibinfo{pages}{1--9}.
\newblock
\urldef\tempurl%
\url{https://doi.org/10.1109/FCCM53951.2022.9786118}
\showDOI{\tempurl}


\bibitem[Ender et~al\mbox{.}(2020)]%
        {ender2020}
\bibfield{author}{\bibinfo{person}{Maik Ender}, \bibinfo{person}{Amir Moradi},
  {and} \bibinfo{person}{Christof Paar}.} \bibinfo{year}{2020}\natexlab{}.
\newblock \showarticletitle{The Unpatchable Silicon: A Full Break of the
  Bitstream Encryption of Xilinx 7-Series {FPGAs}}. In
  \bibinfo{booktitle}{\emph{29th USENIX Security Symposium (USENIX Security
  20)}}. \bibinfo{publisher}{USENIX Association}, \bibinfo{pages}{1803--1819}.
\newblock
\showISBNx{978-1-939133-17-5}
\urldef\tempurl%
\url{https://www.usenix.org/conference/usenixsecurity20/presentation/ender}
\showURL{%
\tempurl}


\bibitem[Ender et~al\mbox{.}(2019)]%
        {ender2019}
\bibfield{author}{\bibinfo{person}{Maik Ender}, \bibinfo{person}{Pawel
  Swierczynski}, \bibinfo{person}{Sebastian Wallat}, \bibinfo{person}{Matthias
  Wilhelm}, \bibinfo{person}{Paul~Martin Knopp}, {and}
  \bibinfo{person}{Christof Paar}.} \bibinfo{year}{2019}\natexlab{}.
\newblock \showarticletitle{Insights into the Mind of a Trojan Designer: The
  Challenge to Integrate a Trojan into the Bitstream}. In
  \bibinfo{booktitle}{\emph{Proceedings of the 24th Asia and South Pacific
  Design Automation Conference}} (Tokyo, Japan).
  \bibinfo{publisher}{Association for Computing Machinery},
  \bibinfo{address}{New York, NY, USA}, \bibinfo{pages}{112–119}.
\newblock
\showISBNx{9781450360074}
\urldef\tempurl%
\url{https://doi.org/10.1145/3287624.3288742}
\showDOI{\tempurl}


\bibitem[Engels et~al\mbox{.}(2023)]%
        {engels2023}
\bibfield{author}{\bibinfo{person}{Susanne Engels}, \bibinfo{person}{Maik
  Ender}, {and} \bibinfo{person}{Christof Paar}.}
  \bibinfo{year}{2023}\natexlab{}.
\newblock \showarticletitle{Targeted Bitstream Fault Fuzzing Accelerating BiFI
  on Large Designs}. In \bibinfo{booktitle}{\emph{2023 IEEE International
  Symposium on Hardware Oriented Security and Trust (HOST)}}.
  \bibinfo{pages}{13--23}.
\newblock
\urldef\tempurl%
\url{https://doi.org/10.1109/HOST55118.2023.10133494}
\showDOI{\tempurl}


\bibitem[F-Secure(2019)]%
        {CVEXilinxUSHack}
\bibfield{author}{\bibinfo{person}{F-Secure}.} \bibinfo{year}{2019}\natexlab{}.
\newblock \bibinfo{title}{{CVE}-2019-5478}.
\newblock
\newblock
\urldef\tempurl%
\url{https://github.com/f-secure-foundry/advisories/blob/master/Security_Advisory-Ref_FSC-HWSEC-VR2019-0001-Xilinx_ZU+-Encrypt_Only_Secure_Boot_bypass.txt}
\showURL{%
Retrieved July 4, 2024 from \tempurl}


\bibitem[f4pga(2017)]%
        {prjxray}
\bibfield{author}{\bibinfo{person}{f4pga}.} \bibinfo{year}{2017}\natexlab{}.
\newblock \bibinfo{title}{Project X-Ray -- {X}ilinx Series 7 Bitstream
  Documentation}.
\newblock
\newblock
\urldef\tempurl%
\url{https://github.com/f4pga/prjxray}
\showURL{%
Retrieved July 4, 2024 from \tempurl}


\bibitem[Graham et~al\mbox{.}(2001)]%
        {graham2001}
\bibfield{author}{\bibinfo{person}{Paul Graham}, \bibinfo{person}{Brent
  Nelson}, {and} \bibinfo{person}{Brad Hutchings}.}
  \bibinfo{year}{2001}\natexlab{}.
\newblock \showarticletitle{Instrumenting Bitstreams for Debugging FPGA
  Circuits}. In \bibinfo{booktitle}{\emph{Proceedings of the the 9th Annual
  IEEE Symposium on Field-Programmable Custom Computing Machines}}.
  \bibinfo{publisher}{IEEE Computer Society}, \bibinfo{address}{USA},
  \bibinfo{pages}{41–50}.
\newblock
\showISBNx{0769526675}


\bibitem[Götte and Scheuermann(2021)]%
        {goette2021}
\bibfield{author}{\bibinfo{person}{Jan~Sebastian Götte} {and}
  \bibinfo{person}{Björn Scheuermann}.} \bibinfo{year}{2021}\natexlab{}.
\newblock \showarticletitle{Can’t Touch This: Inertial HSMs Thwart Advanced
  Physical Attacks}.
\newblock \bibinfo{journal}{\emph{IACR Transactions on Cryptographic Hardware
  and Embedded Systems}} \bibinfo{volume}{2022}, \bibinfo{number}{1}
  (\bibinfo{date}{Nov.} \bibinfo{year}{2021}), \bibinfo{pages}{69–93}.
\newblock
\urldef\tempurl%
\url{https://doi.org/10.46586/tches.v2022.i1.69-93}
\showDOI{\tempurl}


\bibitem[Hazari et~al\mbox{.}(2018)]%
        {hazari2018}
\bibfield{author}{\bibinfo{person}{Noor~Ahmad Hazari}, \bibinfo{person}{Faris
  Alsulami}, {and} \bibinfo{person}{Mohammed Niamat}.}
  \bibinfo{year}{2018}\natexlab{}.
\newblock \showarticletitle{FPGA IP Obfuscation Using Ring Oscillator Physical
  Unclonable Function}. In \bibinfo{booktitle}{\emph{NAECON 2018 - IEEE
  National Aerospace and Electronics Conference}}. \bibinfo{pages}{105--108}.
\newblock
\urldef\tempurl%
\url{https://doi.org/10.1109/NAECON.2018.8556746}
\showDOI{\tempurl}


\bibitem[Hettwer et~al\mbox{.}(2020)]%
        {Hettwer_Leger_Fennes_Gehrer_Gueneysu_2020}
\bibfield{author}{\bibinfo{person}{Benjamin Hettwer},
  \bibinfo{person}{Sebastien Leger}, \bibinfo{person}{Daniel Fennes},
  \bibinfo{person}{Stefan Gehrer}, {and} \bibinfo{person}{Tim Güneysu}.}
  \bibinfo{year}{2020}\natexlab{}.
\newblock \showarticletitle{{Side-Channel Analysis of the Xilinx Zynq
  UltraScale+ Encryption Engine}}.
\newblock \bibinfo{journal}{\emph{IACR Transactions on Cryptographic Hardware
  and Embedded Systems}} \bibinfo{volume}{2021}, \bibinfo{number}{1}
  (\bibinfo{date}{Dec.} \bibinfo{year}{2020}), \bibinfo{pages}{279--304}.
\newblock
\urldef\tempurl%
\url{https://doi.org/10.46586/tches.v2021.i1.279-304}
\showDOI{\tempurl}


\bibitem[Hutchings and Keeley(2014)]%
        {hutchings2014}
\bibfield{author}{\bibinfo{person}{Brad~L. Hutchings} {and}
  \bibinfo{person}{Jared Keeley}.} \bibinfo{year}{2014}\natexlab{}.
\newblock \showarticletitle{Rapid Post-Map Insertion of Embedded Logic
  Analyzers for Xilinx FPGAs}. In \bibinfo{booktitle}{\emph{2014 IEEE 22nd
  Annual International Symposium on Field-Programmable Custom Computing
  Machines}}. \bibinfo{pages}{72--79}.
\newblock
\urldef\tempurl%
\url{https://doi.org/10.1109/FCCM.2014.29}
\showDOI{\tempurl}


\bibitem[Karam et~al\mbox{.}(2016)]%
        {karam2016}
\bibfield{author}{\bibinfo{person}{Robert Karam}, \bibinfo{person}{Tamzidul
  Hoque}, \bibinfo{person}{Sandip Ray}, \bibinfo{person}{Mark Tehranipoor},
  {and} \bibinfo{person}{Swarup Bhunia}.} \bibinfo{year}{2016}\natexlab{}.
\newblock \showarticletitle{Robust bitstream protection in FPGA-based systems
  through low-overhead obfuscation}. In \bibinfo{booktitle}{\emph{2016
  International Conference on ReConFigurable Computing and FPGAs (ReConFig)}}.
  \bibinfo{pages}{1--8}.
\newblock
\urldef\tempurl%
\url{https://doi.org/10.1109/ReConFig.2016.7857187}
\showDOI{\tempurl}


\bibitem[Kashani et~al\mbox{.}(2022)]%
        {kashani2022}
\bibfield{author}{\bibinfo{person}{Sahand Kashani}, \bibinfo{person}{Mahyar
  Emami}, {and} \bibinfo{person}{James~R. Larus}.}
  \bibinfo{year}{2022}\natexlab{}.
\newblock \showarticletitle{Bitfiltrator: A general approach for
  reverse-engineering Xilinx bitstream formats}. In
  \bibinfo{booktitle}{\emph{2022 32nd International Conference on
  Field-Programmable Logic and Applications (FPL)}}. \bibinfo{pages}{01--08}.
\newblock
\urldef\tempurl%
\url{https://doi.org/10.1109/FPL57034.2022.00039}
\showDOI{\tempurl}


\bibitem[Kataria et~al\mbox{.}(2019)]%
        {kataria2019}
\bibfield{author}{\bibinfo{person}{Jatin Kataria}, \bibinfo{person}{Rick
  Housley}, \bibinfo{person}{Joseph Pantoga}, {and} \bibinfo{person}{Ang Cui}.}
  \bibinfo{year}{2019}\natexlab{}.
\newblock \showarticletitle{Defeating Cisco Trust Anchor: A Case-Study of
  Recent Advancements in Direct FPGA Bitstream Manipulation}. In
  \bibinfo{booktitle}{\emph{WOOT @ USENIX Security Symposium}}.
\newblock
\urldef\tempurl%
\url{https://api.semanticscholar.org/CorpusID:201803833}
\showURL{%
\tempurl}


\bibitem[Klix et~al\mbox{.}(2024)]%
        {klix2024}
\bibfield{author}{\bibinfo{person}{Simon Klix}, \bibinfo{person}{Nils
  Albartus}, \bibinfo{person}{Julian Speith}, \bibinfo{person}{Paul Staat},
  \bibinfo{person}{Alice Verstege}, \bibinfo{person}{Annika Wilde},
  \bibinfo{person}{Daniel Lammers}, \bibinfo{person}{Jörn Langheinrich},
  \bibinfo{person}{Christian Kison}, \bibinfo{person}{Sebastian Sester-Wehle},
  \bibinfo{person}{Daniel Holcomb}, {and} \bibinfo{person}{Christof Paar}.}
  \bibinfo{year}{2024}\natexlab{}.
\newblock \bibinfo{title}{Stealing Maggie's Secrets -- On the Challenges of IP
  Theft Through FPGA Reverse Engineering}.
\newblock
\newblock
\showeprint[arxiv]{2312.06195}~[cs.CR]
\urldef\tempurl%
\url{https://arxiv.org/abs/2312.06195}
\showURL{%
\tempurl}


\bibitem[Knittel et~al\mbox{.}(2008)]%
        {knittel2008}
\bibfield{author}{\bibinfo{person}{G. Knittel}, \bibinfo{person}{S. Mayer},
  {and} \bibinfo{person}{C. Rothlaender}.} \bibinfo{year}{2008}\natexlab{}.
\newblock \showarticletitle{Integrating Logic Analyzer Functionality into VHDL
  Designs}. In \bibinfo{booktitle}{\emph{2008 International Conference on
  Reconfigurable Computing and FPGAs}}. \bibinfo{pages}{127--132}.
\newblock
\urldef\tempurl%
\url{https://doi.org/10.1109/ReConFig.2008.77}
\showDOI{\tempurl}


\bibitem[Lavin and Kaviani(2018)]%
        {lavin2018}
\bibfield{author}{\bibinfo{person}{Chris Lavin} {and} \bibinfo{person}{Alireza
  Kaviani}.} \bibinfo{year}{2018}\natexlab{}.
\newblock \showarticletitle{RapidWright: Enabling Custom Crafted
  Implementations for FPGAs}. In \bibinfo{booktitle}{\emph{2018 IEEE 26th
  Annual International Symposium on Field-Programmable Custom Computing
  Machines (FCCM)}}. \bibinfo{pages}{133--140}.
\newblock
\urldef\tempurl%
\url{https://doi.org/10.1109/FCCM.2018.00030}
\showDOI{\tempurl}


\bibitem[Lohrke et~al\mbox{.}(2018)]%
        {lohrke2018}
\bibfield{author}{\bibinfo{person}{Heiko Lohrke}, \bibinfo{person}{Shahin
  Tajik}, \bibinfo{person}{Thilo Krachenfels}, \bibinfo{person}{Christian
  Boit}, {and} \bibinfo{person}{Jean-Pierre Seifert}.}
  \bibinfo{year}{2018}\natexlab{}.
\newblock \showarticletitle{Key Extraction Using Thermal Laser Stimulation: A
  Case Study on Xilinx Ultrascale FPGAs}.
\newblock \bibinfo{journal}{\emph{IACR Transactions on Cryptographic Hardware
  and Embedded Systems}} \bibinfo{volume}{2018}, \bibinfo{number}{3}
  (\bibinfo{date}{Aug.} \bibinfo{year}{2018}), \bibinfo{pages}{573–595}.
\newblock
\urldef\tempurl%
\url{https://doi.org/10.13154/tches.v2018.i3.573-595}
\showDOI{\tempurl}


\bibitem[Mardani~Kamali et~al\mbox{.}(2018)]%
        {kamali2018}
\bibfield{author}{\bibinfo{person}{Hadi Mardani~Kamali}, \bibinfo{person}{Kimia
  Zamiri~Azar}, \bibinfo{person}{Kris Gaj}, \bibinfo{person}{Houman Homayoun},
  {and} \bibinfo{person}{Avesta Sasan}.} \bibinfo{year}{2018}\natexlab{}.
\newblock \showarticletitle{LUT-Lock: A Novel LUT-Based Logic Obfuscation for
  FPGA-Bitstream and ASIC-Hardware Protection}. In
  \bibinfo{booktitle}{\emph{2018 IEEE Computer Society Annual Symposium on VLSI
  (ISVLSI)}}. \bibinfo{pages}{405--410}.
\newblock
\urldef\tempurl%
\url{https://doi.org/10.1109/ISVLSI.2018.00080}
\showDOI{\tempurl}


\bibitem[Moradi et~al\mbox{.}(2013)]%
        {moradi2013}
\bibfield{author}{\bibinfo{person}{Amir Moradi}, \bibinfo{person}{David
  Oswald}, \bibinfo{person}{Christof Paar}, {and} \bibinfo{person}{Pawel
  Swierczynski}.} \bibinfo{year}{2013}\natexlab{}.
\newblock \showarticletitle{Side-Channel Attacks on the Bitstream Encryption
  Mechanism of Altera Stratix II: Facilitating Black-Box Analysis Using
  Software Reverse-Engineering}. In \bibinfo{booktitle}{\emph{Proceedings of
  the ACM/SIGDA International Symposium on Field Programmable Gate Arrays}}
  (Monterey, California, USA). \bibinfo{publisher}{Association for Computing
  Machinery}, \bibinfo{address}{New York, NY, USA}, \bibinfo{pages}{91–100}.
\newblock
\showISBNx{9781450318877}
\urldef\tempurl%
\url{https://doi.org/10.1145/2435264.2435282}
\showDOI{\tempurl}


\bibitem[Moradi and Schneider(2016)]%
        {moradi2016}
\bibfield{author}{\bibinfo{person}{Amir Moradi} {and} \bibinfo{person}{Tobias
  Schneider}.} \bibinfo{year}{2016}\natexlab{}.
\newblock \showarticletitle{Improved Side-Channel Analysis Attacks on Xilinx
  Bitstream Encryption of 5, 6, and 7 Series}. In
  \bibinfo{booktitle}{\emph{Constructive Side-Channel Analysis and Secure
  Design}}. \bibinfo{publisher}{Springer International Publishing},
  \bibinfo{address}{Cham}, \bibinfo{pages}{71--87}.
\newblock
\showISBNx{978-3-319-43283-0}


\bibitem[Moraitis(2023)]%
        {DBLP:journals/access/Moraitis23}
\bibfield{author}{\bibinfo{person}{Michail Moraitis}.}
  \bibinfo{year}{2023}\natexlab{}.
\newblock \showarticletitle{{FPGA} Bitstream Modification: Attacks and
  Countermeasures}.
\newblock \bibinfo{journal}{\emph{{IEEE} Access}}  \bibinfo{volume}{11}
  (\bibinfo{year}{2023}), \bibinfo{pages}{127931--127955}.
\newblock
\urldef\tempurl%
\url{https://doi.org/10.1109/ACCESS.2023.3331507}
\showDOI{\tempurl}


\bibitem[Moraitis and Dubrova(2020)]%
        {DBLP:conf/date/MoraitisD20}
\bibfield{author}{\bibinfo{person}{Michail Moraitis} {and}
  \bibinfo{person}{Elena Dubrova}.} \bibinfo{year}{2020}\natexlab{}.
\newblock \showarticletitle{Bitstream Modification Attack on {SNOW} 3G}. In
  \bibinfo{booktitle}{\emph{2020 Design, Automation {\&} Test in Europe
  Conference {\&} Exhibition, {DATE} 2020, Grenoble, France, March 9-13,
  2020}}. \bibinfo{publisher}{{IEEE}}, \bibinfo{pages}{1275--1278}.
\newblock
\urldef\tempurl%
\url{https://doi.org/10.23919/DATE48585.2020.9116222}
\showDOI{\tempurl}


\bibitem[Ni et~al\mbox{.}(2024)]%
        {DBLP:conf/date/NiK0O24}
\bibfield{author}{\bibinfo{person}{Ziying Ni}, \bibinfo{person}{Ayesha Khalid},
  \bibinfo{person}{Weiqiang Liu}, {and} \bibinfo{person}{M{\'{a}}ire O'Neill}.}
  \bibinfo{year}{2024}\natexlab{}.
\newblock \showarticletitle{Bitstream Fault Injection Attacks on {CRYSTALS}
  Kyber Implementations on FPGAs}. In \bibinfo{booktitle}{\emph{Design,
  Automation {\&} Test in Europe Conference {\&} Exhibition, {DATE} 2024,
  Valencia, Spain, March 25-27, 2024}}. \bibinfo{publisher}{{IEEE}},
  \bibinfo{pages}{1--6}.
\newblock
\urldef\tempurl%
\url{https://ieeexplore.ieee.org/document/10546550}
\showURL{%
\tempurl}


\bibitem[Olney and Karam(2020)]%
        {olney2020}
\bibfield{author}{\bibinfo{person}{Brooks Olney} {and} \bibinfo{person}{Robert
  Karam}.} \bibinfo{year}{2020}\natexlab{}.
\newblock \showarticletitle{WATERMARCH: IP Protection Through Authenticated
  Obfuscation in FPGA Bitstreams}.
\newblock \bibinfo{journal}{\emph{IEEE Embedded Systems Letters}}
  \bibinfo{volume}{PP} (\bibinfo{date}{Aug.} \bibinfo{year}{2020}),
  \bibinfo{pages}{1--1}.
\newblock
\urldef\tempurl%
\url{https://doi.org/10.1109/LES.2020.3015092}
\showDOI{\tempurl}


\bibitem[Poppitz(2006)]%
        {sump2006}
\bibfield{author}{\bibinfo{person}{Michael Poppitz}.}
  \bibinfo{year}{2006}\natexlab{}.
\newblock \bibinfo{title}{SUMP: FPGA Based Logic Analyzer}.
\newblock
\newblock
\urldef\tempurl%
\url{https://www.sump.org/projects/analyzer/}
\showURL{%
Retrieved October 15, 2023 from \tempurl}


\bibitem[Rahman et~al\mbox{.}(2021)]%
        {rahman2021}
\bibfield{author}{\bibinfo{person}{Fahim Rahman}, \bibinfo{person}{Farimah
  Farahmandi}, {and} \bibinfo{person}{Mark Tehranipoor}.}
  \bibinfo{year}{2021}\natexlab{}.
\newblock \showarticletitle{An End-to-End Bitstream Tamper Attack Against
  Flip-Chip FPGAs}.
\newblock \bibinfo{howpublished}{Cryptology ePrint Archive, Paper 2021/1542}.
\newblock \bibinfo{journal}{\emph{IACR Cryptology ePrint Archive}}
  \bibinfo{volume}{2021} (\bibinfo{year}{2021}), \bibinfo{pages}{1542}.
\newblock
\urldef\tempurl%
\url{https://eprint.iacr.org/2021/1542}
\showURL{%
\tempurl}
\newblock
\shownote{\url{https://eprint.iacr.org/2021/1542}}.


\bibitem[Rath(2005)]%
        {openocd}
\bibfield{author}{\bibinfo{person}{Dominic Rath}.}
  \bibinfo{year}{2005}\natexlab{}.
\newblock \bibinfo{title}{OpenOCD -- the Open On-Chip Debugger}.
\newblock
\newblock
\urldef\tempurl%
\url{https://openocd.org}
\showURL{%
Retrieved July 4, 2024 from \tempurl}


\bibitem[sigrok(2010)]%
        {sigrok}
\bibfield{author}{\bibinfo{person}{sigrok}.} \bibinfo{year}{2010}\natexlab{}.
\newblock \bibinfo{title}{Sigrok -- portable, cross-platform,
  Free/Libre/Open-Source signal analysis software suite}.
\newblock
\newblock
\urldef\tempurl%
\url{https://sigrok.org}
\showURL{%
Retrieved July 4, 2024 from \tempurl}


\bibitem[Skorobogatov and Woods(2012)]%
        {skorobogatov2012breakthrough}
\bibfield{author}{\bibinfo{person}{Sergei Skorobogatov} {and}
  \bibinfo{person}{Christopher Woods}.} \bibinfo{year}{2012}\natexlab{}.
\newblock \showarticletitle{{Breakthrough Silicon Scanning Discovers Backdoor
  in Military Chip}}. In \bibinfo{booktitle}{\emph{Cryptographic Hardware and
  Embedded Systems - {CHES} 2012 - 14th International Workshop, Leuven,
  Belgium, September 9-12, 2012. Proceedings}}, Vol.~\bibinfo{volume}{7428}.
  \bibinfo{publisher}{Springer}, \bibinfo{pages}{23--40}.
\newblock
\urldef\tempurl%
\url{https://doi.org/10.1007/978-3-642-33027-8\_2}
\showDOI{\tempurl}


\bibitem[Snyder et~al\mbox{.}(2007)]%
        {verilator2007}
\bibfield{author}{\bibinfo{person}{Wilson Snyder}, \bibinfo{person}{Paul
  Wasson}, {and} \bibinfo{person}{Duane Galbi}.}
  \bibinfo{year}{2007}\natexlab{}.
\newblock \bibinfo{title}{Verilator}.
\newblock
\newblock
\urldef\tempurl%
\url{https://www.veripool.com/verilator}
\showURL{%
Retrieved July 4, 2024 from \tempurl}


\bibitem[Staat et~al\mbox{.}(2022)]%
        {staat2022}
\bibfield{author}{\bibinfo{person}{Paul Staat}, \bibinfo{person}{Johannes
  Tobisch}, \bibinfo{person}{Christian Zenger}, {and} \bibinfo{person}{Christof
  Paar}.} \bibinfo{year}{2022}\natexlab{}.
\newblock \showarticletitle{Anti-Tamper Radio: System-Level Tamper Detection
  for Computing Systems}. In \bibinfo{booktitle}{\emph{2022 IEEE Symposium on
  Security and Privacy (SP)}}. \bibinfo{pages}{1722--1736}.
\newblock
\urldef\tempurl%
\url{https://doi.org/10.1109/SP46214.2022.9833631}
\showDOI{\tempurl}


\bibitem[Swierczynski et~al\mbox{.}(2017)]%
        {swierczynski2017}
\bibfield{author}{\bibinfo{person}{Pawel Swierczynski}, \bibinfo{person}{Georg
  Becker}, \bibinfo{person}{Amir Moradi}, {and} \bibinfo{person}{Christof
  Paar}.} \bibinfo{year}{2017}\natexlab{}.
\newblock \showarticletitle{Bitstream Fault Injections (BiFI) – Automated
  Fault Attacks against SRAM-based FPGAs}.
\newblock \bibinfo{journal}{\emph{IEEE Trans. Comput.}}  \bibinfo{volume}{PP}
  (\bibinfo{date}{Jan.} \bibinfo{year}{2017}), \bibinfo{pages}{1--1}.
\newblock
\urldef\tempurl%
\url{https://doi.org/10.1109/TC.2016.2646367}
\showDOI{\tempurl}


\bibitem[Swierczynski et~al\mbox{.}(2016)]%
        {swierczynski2016}
\bibfield{author}{\bibinfo{person}{Pawel Swierczynski}, \bibinfo{person}{Marc
  Fyrbiak}, \bibinfo{person}{Philipp Koppe}, \bibinfo{person}{Amir Moradi},
  {and} \bibinfo{person}{Christof Paar}.} \bibinfo{year}{2016}\natexlab{}.
\newblock \showarticletitle{Interdiction in practice{\textemdash}Hardware
  Trojan against a high-security {USB} flash drive}.
\newblock \bibinfo{journal}{\emph{Journal of Cryptographic Engineering}}
  \bibinfo{volume}{7}, \bibinfo{number}{3} (\bibinfo{date}{Jun.}
  \bibinfo{year}{2016}), \bibinfo{pages}{199--211}.
\newblock
\urldef\tempurl%
\url{https://doi.org/10.1007/s13389-016-0132-7}
\showDOI{\tempurl}


\bibitem[Swierczynski et~al\mbox{.}(2015)]%
        {swierczynski2015}
\bibfield{author}{\bibinfo{person}{Pawel Swierczynski}, \bibinfo{person}{Marc
  Fyrbiak}, \bibinfo{person}{Philipp Koppe}, {and} \bibinfo{person}{Christof
  Paar}.} \bibinfo{year}{2015}\natexlab{}.
\newblock \showarticletitle{FPGA Trojans Through Detecting and Weakening of
  Cryptographic Primitives}.
\newblock \bibinfo{journal}{\emph{IEEE Transactions on Computer-Aided Design of
  Integrated Circuits and Systems}}  \bibinfo{volume}{34} (\bibinfo{date}{Aug.}
  \bibinfo{year}{2015}), \bibinfo{pages}{1--1}.
\newblock
\urldef\tempurl%
\url{https://doi.org/10.1109/TCAD.2015.2399455}
\showDOI{\tempurl}


\bibitem[Xilinx(2020)]%
        {xilinxchipscope}
\bibfield{author}{\bibinfo{person}{Xilinx}.} \bibinfo{year}{2020}\natexlab{}.
\newblock \bibinfo{title}{Vivado Design Suite User Guide -- Programming and
  Debugging {(UG908)}}.
\newblock
\newblock
\urldef\tempurl%
\url{https://docs.xilinx.com/v/u/2020.1-English/ug908-vivado-programming-debugging}
\showURL{%
Retrieved July 4, 2024 from \tempurl}


\bibitem[Young and Yung(1996)]%
        {young1996}
\bibfield{author}{\bibinfo{person}{Adam Young} {and} \bibinfo{person}{Moti
  Yung}.} \bibinfo{year}{1996}\natexlab{}.
\newblock \showarticletitle{The Dark Side of ``Black-Box'' Cryptography or:
  Should We Trust Capstone?}. In \bibinfo{booktitle}{\emph{Advances in
  Cryptology --- CRYPTO '96}}. \bibinfo{publisher}{Springer Berlin Heidelberg},
  \bibinfo{address}{Berlin, Heidelberg}, \bibinfo{pages}{89--103}.
\newblock
\showISBNx{978-3-540-68697-2}


\bibitem[Yu et~al\mbox{.}(2018)]%
        {yu2018}
\bibfield{author}{\bibinfo{person}{Hoyoung Yu}, \bibinfo{person}{Hansol Lee},
  \bibinfo{person}{Sangil Lee}, \bibinfo{person}{Youngmin Kim}, {and}
  \bibinfo{person}{Hyung-Min Lee}.} \bibinfo{year}{2018}\natexlab{}.
\newblock \showarticletitle{Recent Advances in FPGA Reverse Engineering}.
\newblock \bibinfo{journal}{\emph{Electronics}}  \bibinfo{volume}{7}
  (\bibinfo{date}{Oct.} \bibinfo{year}{2018}), \bibinfo{pages}{246}.
\newblock
\urldef\tempurl%
\url{https://doi.org/10.3390/electronics7100246}
\showDOI{\tempurl}


\bibitem[Zhang et~al\mbox{.}(2019)]%
        {zhang2019}
\bibfield{author}{\bibinfo{person}{Tao Zhang}, \bibinfo{person}{Jian Wang},
  \bibinfo{person}{Shize Guo}, {and} \bibinfo{person}{Zhe Chen}.}
  \bibinfo{year}{2019}\natexlab{}.
\newblock \showarticletitle{A Comprehensive FPGA Reverse Engineering
  Tool-Chain: From Bitstream to RTL Code}.
\newblock \bibinfo{journal}{\emph{IEEE Access}}  \bibinfo{volume}{7}
  (\bibinfo{year}{2019}), \bibinfo{pages}{38379--38389}.
\newblock
\urldef\tempurl%
\url{https://doi.org/10.1109/ACCESS.2019.2901949}
\showDOI{\tempurl}


\end{thebibliography}

\appendix
\newpage
\onecolumn
\section{Test Vectors}
In this appendix we supply \ac{IO} data traces for the original design as well as the Hardware Trojan case studies 2 - 4 that have a different \ac{IO} behavior according to their functionality.%\vspace*{0.2cm}

\subsection{Unmodified Target Application Output}
\label{fpgamod:app:original}
{\small
\begin{tabular}{p{1.8cm}p{2.7cm}l}
    Key:    & $k$ & \texttt{00 01 02 03 04 05 06 07 08 09 0A 0B 0C 0D 0E 0F} \\ \hline
    Input:  & $m$ & \texttt{00 00 00 00 00 00 00 00 00 00 00 00 00 00 00 00} \\ \hline
    Output: & $enc_k(m)$ & \texttt{C6 A1 3B 37 87 8F 5B 82 6F 4F 81 62 A1 C8 D8 79} \\
    & $enc_k(m+1)$ & \texttt{73 46 13 95 95 C0 B4 1E 49 7B BD E3 65 F4 2D 0A} \\
    & $enc_k(m+2)$ & \texttt{49 D6 87 53 99 9B A6 8C E3 89 7A 68 60 81 B0 9D} \\
    & $enc_k(m+3)$ & \texttt{B9 AD 2B 2E 34 6A C2 38 50 5D 36 5E 9C B7 FC 56} \\
    & \dots & \\
\end{tabular}
}

\subsection{Output of Case Study 2}
\label{fpgamod:app:cs2}
{\small
\begin{tabular}{p{1.8cm}p{2.7cm}l}
    Key:        & $k$ & \texttt{00 01 02 03 04 05 06 07 08 09 0A 0B 0C 0D 0E 0F} \\ \hline
    Trojan Key: & $k_{\text{Trojan}}$ & \texttt{00 11 22 33 44 55 66 77 88 99 AA BB CC DD EE FF} \\ \hline
    Input:      & $m$ & \texttt{00 00 00 00 00 00 00 00 00 00 00 00 00 00 00 00} \\ \hline
    Output:     & $enc_{k_{\text{Trojan}}}(k)$ & \texttt{27 9F B7 4A 75 72 13 5E 8F 9B 8E F6 D1 EE E0 03} \\
    & $enc_k(m+1)$ & \texttt{73 46 13 95 95 C0 B4 1E 49 7B BD E3 65 F4 2D 0A} \\
    & $enc_k(m+2)$ & \texttt{49 D6 87 53 99 9B A6 8C E3 89 7A 68 60 81 B0 9D} \\
    & $enc_k(m+3)$ & \texttt{B9 AD 2B 2E 34 6A C2 38 50 5D 36 5E 9C B7 FC 56} \\
    & \dots & \\
\end{tabular}
}

\subsection{Output of Case Study 3}
\label{fpgamod:app:cs3}
{\small
\begin{tabular}{p{1.8cm}p{2.7cm}l}
    Trojan Key: & $k_{\text{Trojan}}$ & \texttt{00 11 22 33 44 55 66 77 88 99 AA BB CC DD EE FF} \\ \hline
    Input:      & $m$ & \texttt{00 00 00 00 00 00 00 00 00 00 00 00 00 00 00 00} \\ \hline
    Output:     & $enc_{k_{\text{Trojan}}}(m)$ & \texttt{FD E4 FB AE 4A 09 E0 20 EF F7 22 96 9F 83 83 2B} \\
    & $enc_{k_{\text{Trojan}}}(m+1)$ & \texttt{84 D4 C9 C0 8B 4F 48 28 61 E3 A9 C6 C3 5B C4 D9} \\
    & $enc_{k_{\text{Trojan}}}(m+2)$ & \texttt{1D F9 27 37 45 13 BF D4 9F 43 6B D7 3F 32 52 85} \\
    & \dots & \\
\end{tabular}
}

\subsection{Output of Case Study 4}
\label{fpgamod:app:cs4}
{\small
\begin{tabular}{p{1.8cm}p{2.7cm}l}
    Key:        & $k$ & \texttt{00 01 02 03 04 05 06 07 08 09 0A 0B 0C 0D 0E 0F} \\ \hline
    Trojan Key: & $k_{\text{Trojan}}$ & \texttt{00 11 22 33 44 55 66 77 88 99 AA BB CC DD EE FF} \\ \hline
    Input:  & $m$ & \texttt{00 00 00 00 00 00 00 00 00 00 00 00 00 00 00 00} \\ \hline
    Output: & $enc_{k_{\text{Trojan}}}(k)$ & \texttt{27 9F B7 4A 75 72 13 5E 8F 9B 8E F6 D1 EE E0 03} \\
    & $enc_k(m)$   & \texttt{C6 A1 3B 37 87 8F 5B 82 6F 4F 81 62 A1 C8 D8 79} \\
    & $enc_k(m+1)$ & \texttt{73 46 13 95 95 C0 B4 1E 49 7B BD E3 65 F4 2D 0A} \\
    & $enc_k(m+2)$ & \texttt{49 D6 87 53 99 9B A6 8C E3 89 7A 68 60 81 B0 9D} \\
    & \dots & \\
\end{tabular}
}

\end{document}